\RequirePackage{fix-cm}
\documentclass[twocolumn]{svjour3}          
\smartqed  
\usepackage{hyperref}
\usepackage{xcolor}
\hypersetup{
    colorlinks,
    linkcolor={red!50!black},
    citecolor={blue!50!black},
    urlcolor={blue!80!black}
}
\usepackage{graphicx}
\usepackage{caption}
\usepackage[caption=false]{subfig}
\usepackage{setspace}
\usepackage[misc,geometry]{ifsym}
\usepackage{graphicx}
\usepackage{enumitem}					
\usepackage[english]{babel}
\usepackage{amsmath} 
\usepackage{array}
\newcolumntype{?}{!{\vrule width 1.1pt}}
\usepackage[english]{babel}
\usepackage{blindtext}
\usepackage{mathtools,cuted}
\usepackage{adjustbox}
\usepackage{booktabs}
\usepackage{seqsplit}
\usepackage{etoolbox}
\newcommand*{\affaddr}[1]{#1} 
\newcommand*{\affmark}[1][*]{\textsuperscript{#1}}
\newcommand{\sbt}{\,\begin{picture}(-1,1)(-0.25,-1)\circle*{2}\end{picture}\ }

\usepackage{amssymb}
\usepackage{multirow,makecell}

\begin{document}

\title{Isogeometric Configuration Design Optimization of Three-dimensional Curved Beam Structures for Maximal Fundamental Frequency
}

\titlerunning{Configuration Design Optimization for Maximal Fundamental Frequency}        

\author{Myung-Jin Choi\protect\affmark[1,2] \and Jae-Hyun Kim\affmark[1] \and Bonyong Koo\affmark[2] \and Seonho Cho\affmark[1]}

\authorrunning{M.-J. Choi et al.} 

\institute{
              Corresponding author (S. Cho)\\
              Tel.: +82 2 880 7322\\
              Fax: +82 2 888 9298\\
              \email{secho@snu.ac.kr}\\
\affaddr{\affmark[1]Department of Naval Architecture and Ocean Engineering, Seoul National University, Seoul, Korea}\\
\affaddr{\affmark[2]School of Mechanical Convergence System Engineering, Kunsan National University, Gunsan, Korea}\\
\affaddr{This document is the personal version of an article whose final publication is available at \url{https://doi.org/10.1007/s00158-020-02803-0}.}}

\date{Published in \textit{Structural and Multidisciplinary Optimization}, DOI: \url{https://doi.org/10.1007/s00158-020-02803-0}\\Received: / Accepted: date}

\maketitle

\begin{abstract}
This paper presents a configuration design optimization method for three-dimensional curved beam built-up structures having maximized fundamental eigenfrequency. We develop the method of computation of design velocity field and optimal design of beam structures constrained on a curved surface, where both designs of the embedded beams and the curved surface are simultaneously varied during the optimal design process. A shear-deformable beam model is used in the response analyses of structural vibrations within an isogeometric framework using the NURBS basis functions. An analytical design sensitivity expression of repeated eigenvalues is derived. The developed method is demonstrated through several illustrative examples.


\end{abstract}

\section{Introduction}
A structural design to have eigenfrequencies as far away as possible from excitation frequencies has been regarded as a significant process to avoid resonance and assure structural stability. There have been lots of researches to develop effective and efficient design methods to increase the structural fundamental frequency. Shi et al. \cite{shi2017fundamental} presented a parameter-free approach for a design optimization of orthotropic shells that maximizes fundamental frequencies, where a shape variation was generated by a displacement field due to an external force equivalent to negative shape gradient function. Hu and Tsai \cite{hu1999maximization} found optimal fiber orientation angles to maximize fundamental frequencies of fiber-reinforced laminated cylindrical shells. Blom et al. \cite{blom2008design} optimized the fiber orientation on the surface of a conical shell for the maximization of the fundamental frequency under manufacturing constraints on the curvature of fiber paths. Silva and Nicoletti \cite{da2017optimization} sought an optimal shape design of a beam for desired natural frequencies by a linear combination of mode shapes whose coefficients are determined by an optimization algorithm. Wang et al. \cite{wang2006maximizing} derived an analytic solution of minimum support stiffness for maximizing a natural frequency of a beam. Zhu and Zhang \cite{jihong2006maximization} proposed an optimal support layout for maximizing natural frequency using a topology optimization where stiffness values of support springs are considered as design variables. Allaire and Jouve \cite{allaire2005level} utilized a level-set method to design two- and three-dimensional structures having maximal fundamental frequencies. Picelli et al. \cite{picelli2015evolutionary} applied an evolutionary topology optimization method to maximize the first natural frequency in free-vibration problems considering acoustic-structure interaction. Zuo et al. \cite{zuo2011fast} employed a reanalysis method for eigenvalue problem to reduce computation cost during a design optimization by a genetic algorithm for minimizing the weight of structures with frequency constraints. In the design sensitivity analysis (DSA) of eigenvalue problems, a technical difficulty may arise due to the lack of the usual differentiability in repeated eigenvalues. Seyranian et al. \cite{seyranian1994multiple} presented a perturbation method to derive design sensitivities of repeated eigenvalues, where it was shown that sensitivities of repeated eigenvalues correspond to solutions of a sub-eigenvalue problem. This method was used by Du and Olhoff \cite{du2007topological} for topology optimizations in order to maximize specific eigenfrequencies and gaps between two consecutive eigenfrequencies. It was shown, in Olhoff et al. \cite{olhoff2012optimum}, that the separation of adjacent eigenfrequencies is closely related with the generation of band gap property. \textit{In this paper, we present a gradient-based configuration design optimization method for three-dimensional beam built-up structures having maximal fundamental frequencies, using an analytical configuration DSA for repeated as well as simple eigenvalues.}

Isogeometric analysis (IGA) suggested by Hughes et al. \cite{hughes2005isogeometric} fundamentally pursues a seamless integration of computer-aided design (CAD) and analysis. Especially, within the IGA framework, shape design, and sizing design parameters like beam cross-section thickness and composite fiber orientation angle have higher regularity due to NURBS basis functions than the conventional finite element analysis (FEA) model. Taheri and Hassani \cite{taheri2014simultaneous} combined shape and fiber orientation design optimizations for functionally graded structures to have optimal eigenfrequencies. Nagy et al. \cite{nagy2011isogeometric} performed a sizing and configuration design optimization for a fundamental frequency maximization of planar Euler beams. They also presented several types of manufacturing constraints to avoid the abrupt change of cross-section thickness and neutral axis curvature. Liu et al. \cite{liu2019smooth} presented a sizing optimal design of planar Timoshenko beams, where the maximum rate of change of cross-section thickness was constrained, and the constraints were aggregated by the K-S constraint scheme. These previous works on the isogeometric optimal structural design for maximizing the fundamental frequency using a beam element were limited to the design of two-dimensional structure and single patch model. \textit{In this paper, we deal with configuration design of more practical three-dimensional built-up structures.}

In this work, we particularly focus on the computation of design velocity field and optimal design of beam structures constrained in curved surfaces. Structures geometrically constrained on curved domains have been widely utilized in many engineering applications including substructures for offshore oil and wind turbine platforms. This kind of design problem has equality constraints restricting their position on a specified curved domain, and several researches regarding the constrained optimal design problem have been reported recently. Liu et al. \cite{liu2017additive} performed topology optimization by combining the moving morphable component (MMC) method and coordinate transformations to architect structures constrained in planar curved domains. Choi and Cho \cite{choi2018constrained} presented a constrained isogeometric design optimization method for beam structures located on a specified curved surface, where the free-form deformation (FFD) method was used to locate curves exactly on the surface, and then the curves are interpolated to determine control point positions. Choi and Cho \cite{choi2019isogeometric} presented the construction of initial orthonormal base vectors along a spatial curve embedded in a curved domain by using surface convected frames as reference frames in the smallest rotation (SR) method. However, the previous works \cite{choi2018constrained} and \cite{choi2019isogeometric} assumed that the surfaces where the curves are embedded do not have design dependence, that is, beam structures are designed on a fixed curved surface. \textit{In this paper, extending those previous works, we additionally consider configuration design of the curved surfaces where beam structures are embedded, so that more design degrees-of-freedom are furnished during the constrained optimal design process.}\newline
\indent This article is organized as follows. In section \ref{ig_vib_beam}, we explain an isogeometric analysis of structural vibrations in shear-deformable beam structures. In section \ref{ig_vib_config_dsa}, we explain a configuration DSA of simple and repeated eigenvalues for the beam structures, where a design velocity computation method for constrained beam structures is detailed. In section \ref{ig_vib_num_ex}, the developed optimization method is demonstrated through several numerical examples for maximizations of fundamental frequencies.

\section{Isogeometric analysis of vibrations in shear-deformable beam structures}
\label{ig_vib_beam}
\subsection{NURBS curve geometry}
In this section, we briefly review a NURBS curve exploited to describe a beam neutral axis geometry. A set of knots in one-dimensional space is denoted by
\begin{equation} \label{ig_vib_set_knot}
{\boldsymbol{\xi }} = \left\{ {{\xi _1},{\xi _2}, \cdots ,{\xi _{n + p + 1}}} \right\},
\end{equation}
where $p$ and $n$ are respectively the order of basis function, and the number of control points. The B-spline basis functions are obtained, in a recursive manner, as
\begin{equation} \label{ig_vib_bsp_base0}
N_I^0(\xi ) = \left\{ {\begin{array}{*{20}{c}}
1&{{\rm{    if }}\,\,\,\,{\xi _I} \le \xi  < {\xi _{I + 1}}}\\
0&{{\rm{otherwise  }}}
\end{array}} \right.,\;{\rm{   (}}p = 0{\rm{)}},
\end{equation}
and
\begin{multline} \label{ig_vib_bsp_base}
N_I^p(\xi ) = \frac{{\xi  - {\xi _I}}}{{{\xi _{I + p}} - {\xi _I}}}N_I^{p - 1}(\xi ) + \frac{{{\xi _{I + p + 1}} - \xi }}{{{\xi _{I + p + 1}} - {\xi _{I + 1}}}}N_{I + 1}^{p - 1}(\xi ), \\
\;(p = 1,2,3, \ldots ).
\end{multline}
From the B-spline basis $N_I^p(\xi )$ and the weight $w_I$, the NURBS basis function ${W_I}(\xi )$ is defined by
\begin{equation} \label{ig_vib_nurbs_base}
{W_I}(\xi ) \equiv \frac{{N_I^p(\xi ){w_I}}}{{\sum\limits_{J = 1}^n {N_J^p(\xi ){w_J}} }}.
\end{equation}
The NURBS basis function of order $p$ has $C^{p-k}$ continuity at a knot multiplicity $k$. A curve geometry is described by a linear combination of NURBS basis functions and control point positions as
\begin{equation} \label{ig_vib_nurbs_curve}
{\bf{C}}(\xi ) = \sum\limits_{I = 1}^n {{W_I}(\xi ){{\bf{B}}_I}}.
\end{equation}
More details of the NURBS geometry can be found in \cite{piegl2012nurbs}.

\subsection{Linear kinematics and equilibrium equations}
\begin{figure}
\centering
  \includegraphics[width=0.4\textwidth]{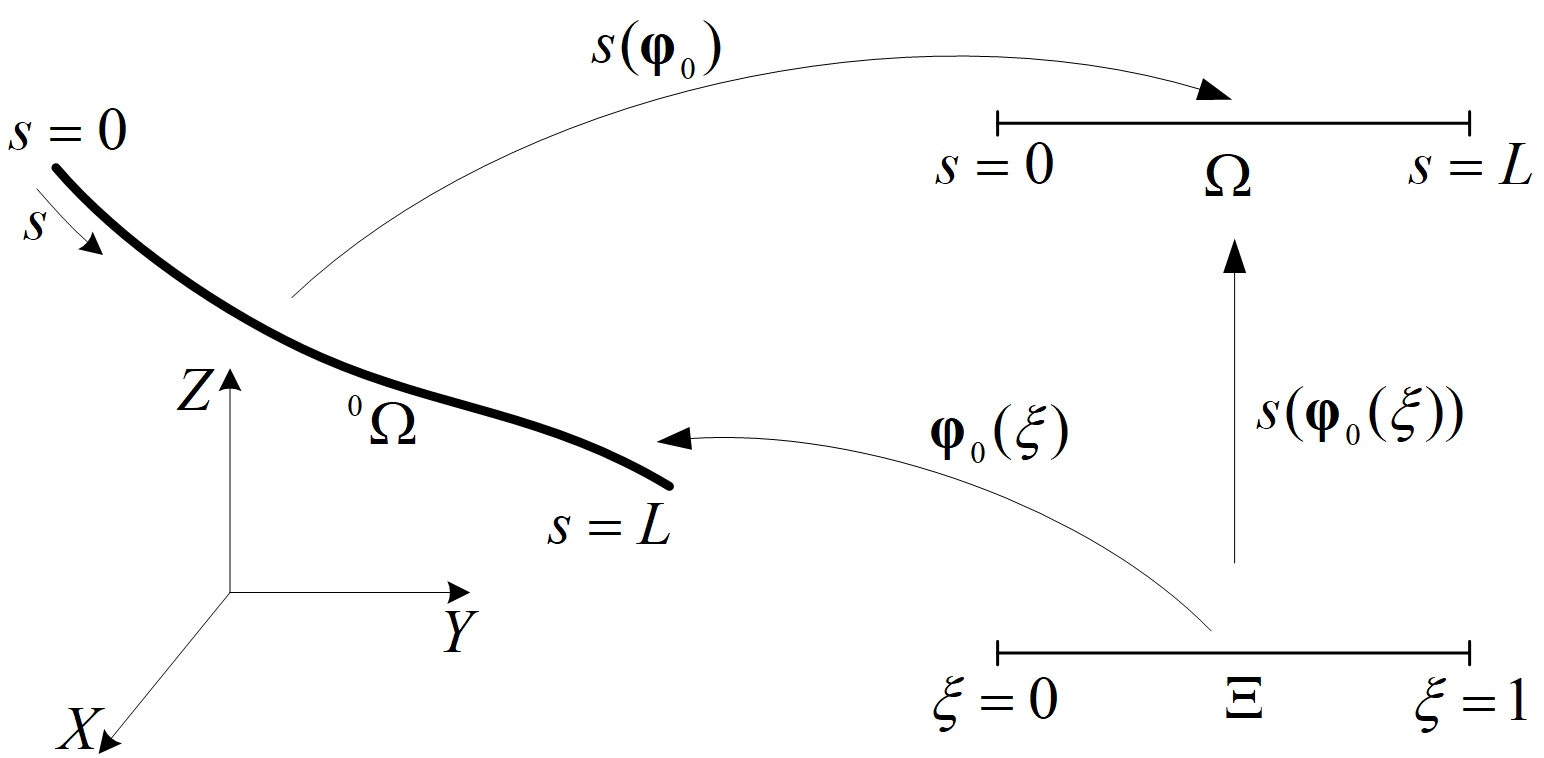}
\caption{Parameterization of a beam neutral axis \cite{choi2019isogeometric,choi2016isogeometric}}
\label{ig_vib_map}       
\end{figure}

\begin{figure}
\centering
  \includegraphics[width=0.4\textwidth]{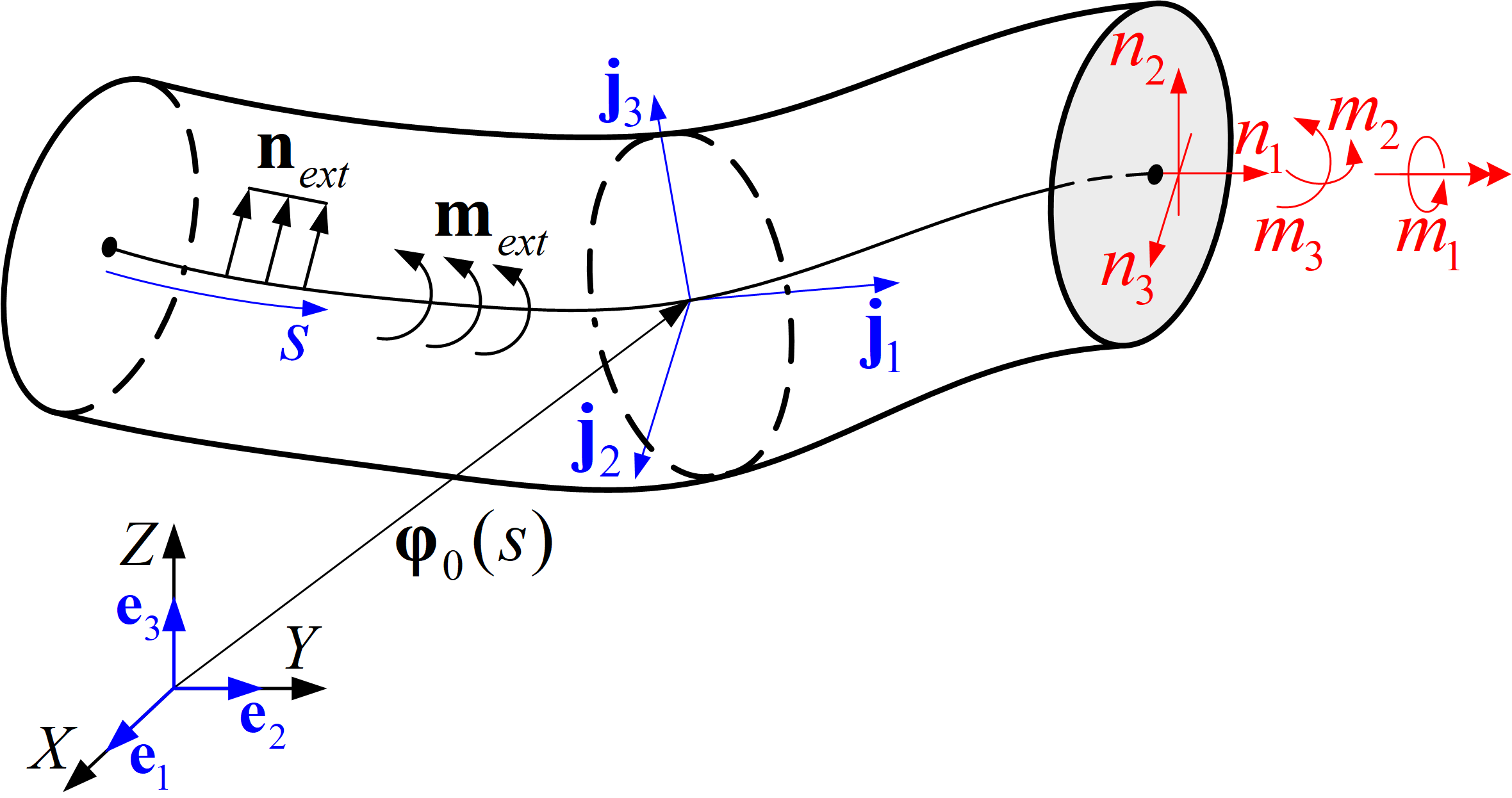}
\caption{Problem definition of a spatial rod: neutral axis defined by general curve ${{\boldsymbol{\varphi }}_0}(s)$ }
\label{ig_vib_prob}       
\end{figure}

Consider an initial beam neutral axis ${{\boldsymbol{\varphi }}_0}(\xi ) \in {{\bf{R}}^3}$ characterized by a NURBS curve of Eq.\,(\ref{ig_vib_nurbs_curve}). We define an arc-length parameter $s \in \Omega  \equiv \left[ {0,L} \right]$ that parameterizes the beam neutral axis with the initial (undeformed) length $L$ in a way that \cite{choi2016isogeometric}
\begin{equation} \label{ig_vib_alen_param}
s({{\boldsymbol{\varphi }}_0}(\xi )) = \int_0^{{\xi _0} = \xi } {\left\| {\frac{{\partial {{\boldsymbol{\varphi }}_0}({\xi _0})}}{{\partial {\xi _0}}}} \right\|d{\xi _0}}  \equiv s(\xi ),
\end{equation}
which relates parametric domains of the arc-length parameter ($\Omega $) and NURBS parametric coordinate ($\Xi $) through the physical domain ${}^0\Omega $, as depicted in Fig. \ref{ig_vib_map}. The Jacobian of the mapping is defined as \cite{choi2016isogeometric}
\begin{equation} \label{ig_vib_curve_jcb}
{J_c} \equiv \frac{{\partial s}}{{\partial \xi }} = \left\| {\frac{{\partial {{\boldsymbol{\varphi }}_0}(\xi )}}{{\partial \xi }}} \right\|.
\end{equation}
Hereafter, we often denote ${{\boldsymbol{\varphi }}_0} \equiv {{\boldsymbol{\varphi }}_0}(s)$ for a brief expression. A cross-section orientation is described by an orthonormal frame with basis ${{\bf{j}}_1}(s),{{\bf{j}}_2}(s),{{\bf{j}}_3}(s) \in {{\bf{R}}^3}$, where the unit tangent vector is obtained by ${{\bf{j}}_1}(s) \equiv {{\boldsymbol{\varphi }}_{0,s}}$ due to the arc-length parameterization, and assumed to be orthogonal to the cross-section (see Fig. \ref{ig_vib_prob} for an illustration). ${(\bullet)_{,s}}$ denotes the partial differentiation with respect to the arc-length parameter. The global Cartesian coordinate system is denoted by $X-Y-Z$ with base vectors ${{\bf{e}}_1},{{\bf{e}}_2},{{\bf{e}}_3} \in {{\bf{R}}^3}$.
In this paper, we consider a structural dynamics with small (infinitesimal) amplitudes, where a linearized kinematics consistently derived from a general nonlinear formulation of \cite{simo1985finite} is utilized. The (material form) axial-shear and bending-torsional strain measures derived in \cite{simo1986three} are evaluated at an undeformed state, which respectively result in \cite{choi2019optimal}
\begin{equation} \label{ig_vib_strn_gamma}
{\bf{\Gamma }}(s,t) \equiv {{\bf{\Lambda }}_0}{(s)^T}\left\{ {\frac{\partial }{{\partial s}}{{\bf{z}}_t} + {{\bf{j}}_1}(s) \times {{\boldsymbol{\theta }}_t}} \right\},
\end{equation}
and
\begin{equation} \label{ig_vib_strn_omg}
{\bf{\Omega }}(s,t) \equiv {{\bf{\Lambda }}_0}{(s)^T}\frac{{\partial {{\boldsymbol{\theta }}_t}}}{{\partial s}}.
\end{equation}
$t \ge 0$ represents the time, and ${{\bf{z}}_t} \equiv {\bf{z}}(s,t)$ denotes the displacement of the neutral axis position, and ${{\boldsymbol{\theta }}_t} \equiv {\boldsymbol{\theta }}(s,t)$ denotes the (infinitesimal) rotation vector that describes the rigid rotation of cross-section. ${{\bf{\Lambda }}_0}(s) \equiv \left[ {{{\bf{j}}_1}(s),{{\bf{j}}_2}(s),{{\bf{j}}_3}(s)} \right]$ defines an orthogonal transformation matrix such that
\begin{equation} \label{ig_vib_jtransform}
{{\bf{j}}_I}(s) = {{\bf{\Lambda }}_0}(s){{\bf{e}}_I},\,\,I=1,2,3.
\end{equation}
We define an elastic strain energy by \cite{choi2019optimal}
\begin{equation} \label{ig_vib_elas_strn_e}
U \equiv \frac{1}{2}\int_\Omega  {\left\{ {{\bf{\Gamma }}(s,t) \cdot {\bf{N}}(s,t) + {\bf{\Omega }}(s,t) \cdot {\bf{M}}(s,t)} \right\}ds},
\end{equation}
where ${\bf{N}}(s,t)$ and ${\bf{M}}(s,t)$ are the material form resultant force and moment vectors over the cross-section at $s \in \Omega $, which are related to corresponding spatial forms through
\begin{equation} \label{ig_vib_nm_transform}
\left. \begin{array}{lcl}
{\bf{n}}(s,t) &=& {{\bf{\Lambda }}_0}(s){\bf{N}}(s,t)\\
{\bf{m}}(s,t) &=& {{\bf{\Lambda }}_0}(s){\bf{M}}(s,t)
\end{array} \right\}.
\end{equation}
As shown in Fig. \ref{ig_vib_prob}, the (spatial form) resultant force and moment can be respectively resolved in the orthonormal basis as ${\bf{n}}(s,t) = {n_I}(s,t){{\bf{j}}_I}(s)$ and ${\bf{m}}(s,t) = {m_I}(s,t){{\bf{j}}_I}(s)$ where the repeated index $I$ implies the summation over the components 1, 2, and 3. We employ a linear elastic constitutive relation as 
\begin{equation} \label{ig_vib_crel}
\left. \begin{array}{lcl}
{\bf{N}}(s,t) &=& {{\bf{C}}_F}{\bf{\Gamma }}(s,t)\\
{\bf{M}}(s,t) &=& {{\bf{C}}_M}{\bf{\Omega }}(s,t)
\end{array} \right\},
\end{equation}
and the constitutive tensors ${{\bf{C}}_F}$ and ${{\bf{C}}_M}$ are defined by
\begin{equation} \label{ig_vib_cmat}
\left. \begin{array}{lcl}
{{\bf{C}}_F} &=& diag{[EA,G{A_2},G{A_3}]_{\left\{ {{{\bf{e}}_I}} \right\}}}\\
{{\bf{C}}_M} &=& diag{[G{I_p},E{I_2},E{I_3}]_{\left\{ {{{\bf{e}}_I}} \right\}}}
\end{array} \right\},
\end{equation}
where $diag[a,b,c]$ represents a diagonal matrix of components $a$, $b$, and $c$. ${(\bullet)_{\left\{ {{{\bf{e}}_I}} \right\}}}$ represents the tensor is represented in the global basis $\left\{ {{{\bf{e}}_1},{{\bf{e}}_2},{{\bf{e}}_3}} \right\}$. $A$ represents the cross-sectional area, and ${A_2} \equiv {k_2}A$, ${A_3} \equiv {k_3}A$ where $k_2$ and $k_3$ denote the shear correction factors. $I_p$ denotes the polar moment of inertia, and $I_2$ and $I_3$ mean the second moments of inertia. $E$ and $G$ denote Young's modulus and shear modulus, respectively. A kinetic energy is defined by \cite{choi2019optimal}
\begin{equation} \label{ig_vib_kin_e}
T \equiv \frac{1}{2}\int_\Omega  {\left\{ {\rho A{{\bf{z}}_{t,t}} \cdot {{\bf{z}}_{t,t}} + {{\boldsymbol{\theta }}_{t,t}} \cdot \left( {{{\bf{I}}_\rho }{{\boldsymbol{\theta }}_{t,t}}} \right)} \right\}ds},
\end{equation}
where ${(\bullet)_{,t}}$ denotes the partial differentiation with respect to the time, and $\rho $ denotes the mass density. It is assumed that the initial local orthonormal basis $\left\{ {{{\bf{j}}_1},{{\bf{j}}_2},{{\bf{j}}_3}} \right\}$ is directed along the principal axes of inertia of the cross-section; thus, the inertia tensor can be expressed by a diagonal matrix in the local basis, and represented in the global basis through the orthogonal transformation of Eq. (\ref{ig_vib_jtransform}), as 
\begin{align} \label{ig_vib_inert_tensor}
{{{\bf{I}}_\rho }}&{ \equiv diag{{[\rho {I_p},\rho {I_2},\rho {I_3}]}_{\left\{ {{{\bf{j}}_I}} \right\}}}}\nonumber\\
{}&{ = {{\bf{\Lambda }}_0}diag{{[\rho {I_p},\rho {I_2},\rho {I_3}]}_{\left\{ {{{\bf{e}}_I}} \right\}}}{{\bf{\Lambda }}_0}^T,} 
\end{align}
where ${(\bullet)_{\left\{ {{{\bf{j}}_I}} \right\}}}$ represents the corresponding tensor is represented in the bases of $\left\{ {{{\bf{j}}_1},{{\bf{j}}_2},{{\bf{j}}_3}} \right\}$. A work done by external loads is expressed as \cite{choi2019optimal}
\begin{equation} \label{ig_vib_work_ext}
W \equiv \int_\Omega  {\left\{ {{{\bf{n}}_{\text{ext}}}(s) \cdot {\bf{z}}(s,t) + {{\bf{m}}_{\text{ext}}}(s) \cdot {\boldsymbol{\theta }}(s,t)} \right\}ds},
\end{equation}
where ${{\bf{n}}_{\text{ext}}}(s)$ and ${{\bf{m}}_{\text{ext}}}(s)$ denote the distributed external force and moment, respectively. In order to derive equilibrium equations from time ${t_1}$ to time ${t_2}$, the Hamilton's principle is employed as \cite{goldstein2002classical} 
\begin{equation} \label{ig_vib_var_hamilton}
\delta \int_{{t_1}}^{{t_2}} {\left( {U - T - W} \right)} dt = 0,
\end{equation}
where $\delta (\bullet)$ defines the first variation. Substituting Eqs. (\ref{ig_vib_elas_strn_e}), (\ref{ig_vib_kin_e}), and (\ref{ig_vib_work_ext}) into Eq. (\ref{ig_vib_var_hamilton}), and applying the integration by parts lead to the following linear and angular momentum balance equations \cite{choi2019optimal}
\begin{equation} \label{ig_vib_gov_eq}
\left. \begin{array}{lcl}
{{\bf{n}}_{t,s}} + {{\bf{n}}_{\text{ext}}} &=& \rho A{{\bf{z}}_{t,tt}}\\
{{\bf{m}}_{t,s}} + {{\bf{j}}_1} \times {{\bf{n}}_t} + {{\bf{m}}_{\text{ext}}} &=& {{\bf{I}}_\rho }{{\boldsymbol{\theta }}_{t,tt}}
\end{array} \right\},
\end{equation}
where ${{\bf{n}}_t} \equiv {\bf{n}}(s,t)$, ${{\bf{m}}_t} \equiv {\bf{m}}(s,t)$, and ${(\bullet)_{,tt}}$ denotes the second-order partial derivative with respect to the time.

\subsection{Variational formulation}
We assume a time-harmonic solution such as
\begin{equation} \label{ig_vib_harmonic_sol}
\left. \begin{array}{lcl}
{\bf{z}}(s,t) &=& {\bf{z}}(s){e^{ - i\omega t}}\\
{\boldsymbol{\theta }}(s,t) &=& {\boldsymbol{\theta }}(s){e^{ - i\omega t}}
\end{array} \right\},
\end{equation}
where $i \equiv \sqrt { - 1} $, and $\omega $ denotes an angular frequency. From the governing equations of Eq.\,(\ref{ig_vib_gov_eq}), using the principle of virtual work and substituting Eq. (\ref{ig_vib_harmonic_sol}), we obtain the following generalized eigenvalue problem. 
\begin{equation} \label{ig_vib_eval_prob}
a({\boldsymbol{\eta }},{\boldsymbol{\bar \eta }}) = \zeta d({\boldsymbol{\eta }},{\boldsymbol{\bar \eta }}),\,\,\forall {\boldsymbol{\bar \eta }} \in \bar Z,
\end{equation}
where $\zeta  \equiv {\omega ^2}$ denotes the eigenvalue associated with the eigenfunction ${\boldsymbol{\eta }} \equiv ({\bf{z}}(s),{\boldsymbol{\theta }}(s))$. $\overline {(\bullet)}  \equiv \delta (\bullet)$ denotes the first variation of 
$(\bullet)$ or the corresponding virtual quantity, and ${\boldsymbol{\bar \eta }} \equiv ({\bf{\bar z}}(s),{\boldsymbol{\bar \theta }}(s))$. The strain energy and kinetic energy bilinear forms are respectively defined as \cite{choi2019optimal}
\begin{equation} \label{ig_vib_strn_e_form}
a({\boldsymbol{\eta }},{\boldsymbol{\bar \eta }}) \equiv \int_\Omega  {\left\{ {\begin{array}{*{20}{c}}
{{\bf{\bar \Gamma }}}\\
{{\bf{\bar \Omega }}}
\end{array}} \right\} \cdot \left\{ {\begin{array}{*{20}{c}}
{\bf{N}}\\
{\bf{M}}
\end{array}} \right\}ds},
\end{equation}
and
\begin{equation} \label{ig_vib_kin_e_form}
d({\boldsymbol{\eta }},{\boldsymbol{\bar \eta }}) \equiv \int_\Omega  {{{\left\{ {\begin{array}{*{20}{c}}
{{\bf{\bar z}}}\\
{{\boldsymbol{\bar \theta }}}
\end{array}} \right\}}^T}\left[ {\begin{array}{*{20}{c}}
{\rho A{\bf{I}}}&{{{\bf{0}}_{3 \times 3}}}\\
{{{\bf{0}}_{3 \times 3}}}&{{{\bf{I}}_\rho }}
\end{array}} \right]\left\{ {\begin{array}{*{20}{c}}
{\bf{z}}\\
{\boldsymbol{\theta }}
\end{array}} \right\}ds},
\end{equation}
where ${{\bf{0}}_{3 \times 3}} \in {{\bf{R}}^{3 \times 3}}$ and ${\bf{I}} \in {{\bf{R}}^{3 \times 3}}$ represent the null and identity matrices, respectively. It is assumed throughout this paper that the eigenfunctions are orthonormalized with respect to the bilinear form $d(\bullet,\bullet)$ such that 
\begin{equation} \label{ig_vib_normalize}
d({\boldsymbol{\eta }},{\boldsymbol{\eta }}) = 1.
\end{equation}
The solution space $\bar Z$ is expressed, considering a homogeneous boundary condition, as
\begin{equation} \label{ig_vib_var_space}
\bar Z = \left\{ {{\boldsymbol{\bar \eta }} \in {H^1} \times {H^1}:{\bf{\bar z}}(0) = {\boldsymbol{\bar \theta }}(0) = {\bf{\bar z}}(L) = {\boldsymbol{\bar \theta }}(L) = {\bf{0}}} \right\},
\end{equation}
where $H^1$ defines the Sobolev space of order 1. The static (material form) axial-shear and bending-torsional strain measures are respectively obtained from Eqs.\,(\ref{ig_vib_strn_gamma}) and (\ref{ig_vib_strn_omg}), as
\begin{equation} \label{ig_vib_lin_strn}
\left. \begin{array}{lcl}
{\bf{\Gamma }}({\boldsymbol{\eta }}) &\equiv& {{\bf{\Lambda }}_0}^T\left( {{{\bf{z}}_{,s}} + {{\bf{j}}_1} \times {\boldsymbol{\theta }}} \right)\\
{\bf{\Omega }}({\boldsymbol{\theta }}) &\equiv& {{\bf{\Lambda }}_0}^T{{\boldsymbol{\theta }}_{,s}}
\end{array} \right\},
\end{equation}
and ${\bf{\bar \Gamma }} \equiv {\bf{\Gamma }}({\boldsymbol{\bar \eta }})$ and ${\bf{\bar \Omega }} \equiv {\bf{\Omega }}({\boldsymbol{\bar \theta }})$ represent the virtual strain measures. Eq.\,(\ref{ig_vib_lin_strn}) can be rewritten in a compact form as
\begin{equation} \label{ig_vib_lin_strn_compact}
\left\{ {\begin{array}{*{20}{c}}
{{\bf{\Gamma }}({\boldsymbol{\eta }})}\\
{{\bf{\Omega }}({\boldsymbol{\theta }})}
\end{array}} \right\} = {{\bf{\Pi }}_0}^T{{\bf{\Xi }}_0}^T\left\{ {\begin{array}{*{20}{c}}
{\bf{z}}\\
{\boldsymbol{\theta }}
\end{array}} \right\},
\end{equation}
using the following matrix operators
\begin{equation} \label{ig_vib_mat_pi0_xi0}
{{\bf{\Pi }}_0} \equiv \left[ {\begin{array}{*{20}{c}}
{{{\bf{\Lambda }}_0}}&{{{\bf{0}}_{3 \times 3}}}\\
{{{\bf{0}}_{3 \times 3}}}&{{{\bf{\Lambda }}_0}}
\end{array}} \right]\,\,\,\,\rm{and}\,\,\,\,
{{\bf{\Xi }}_0} \equiv \left[ {\begin{array}{*{20}{c}}
{\frac{d}{{ds}}{\bf{I}}}&{{{\bf{0}}_{3 \times 3}}}\\
{ - \left[ {{{\bf{j}}_1} \times } \right]}&{\frac{d}{{ds}}{\bf{I}}}
\end{array}} \right],
\end{equation}
where $\left[ {(\bullet) \times } \right]$ denotes a skew-symmetric matrix associated with the dual vector $(\bullet) \in {{\bf{R}}^3}$.

\subsection{Isogeometric discretization using the NURBS basis}
In the IGA framework, the response field is approximated by the same basis functions of NURBS as employed to represent the geometry in CAD. The displacement and rotation amplitudes are approximated by the NURBS basis function, as
\begin{equation} \label{ig_vib_disp_rot_amp_disc}
{{\bf{z}}^h} = \sum\limits_{N = 1}^n {{W_N}(\xi )} {{\bf{y}}_N}\,\,\,\,{\rm{and}}\,\,\,\,{{\boldsymbol{\theta }}^h} = \sum\limits_{N = 1}^n {{W_N}(\xi )} {{\boldsymbol{\theta }}_N},
\end{equation}
where ${{\bf{y}}_N}$ and ${{\boldsymbol{\theta }}_N}$ are the coefficients associated with the $N$-th control point. In the same manner, the variation of the displacement and rotation amplitudes are approximated by
\begin{equation} \label{ig_vib_disp_rot_amp_vir_disc}
{{\bf{\bar z}}^h} = \sum\limits_{N = 1}^n {{W_N}(\xi )} {{\bf{\bar y}}_N}\,\,\,\,{\rm{and}}\,\,\,\,{{\boldsymbol{\bar \theta }}^h} = \sum\limits_{N = 1}^n {{W_N}(\xi )} {{\boldsymbol{\bar \theta }}_N},
\end{equation}
where ${{\bf{\bar y}}_N}$ and ${{\boldsymbol{\bar \theta }}_N}$ are the $N$-th coefficients. Substituting Eq.\,(\ref{ig_vib_disp_rot_amp_disc}) into Eq.\,(\ref{ig_vib_lin_strn_compact}), we have
\begin{equation} \label{ig_vib_strain_disc_compact}
\left\{ {\begin{array}{*{20}{c}}
{{\bf{\Gamma }}({{\bf{\eta }}^h})}\\
{{\bf{\Omega }}({{\bf{\theta }}^h})}
\end{array}} \right\} = {{\bf{\Pi }}_0}^T{{\bf{\Xi }}_{0N}^h}{{\bf{d}}_N}
\end{equation}
where ${{\bf{d}}_N} \equiv {[{{\bf{y}}_N}^T,{{\boldsymbol{\theta }}_N}^T]^T}$, and we use
\begin{equation} \label{ig_vib_mat_psi_disc}
{\bf{\Xi }}_{0N}^h \equiv \left[ {\begin{array}{*{20}{c}}
{{W_{N,s}}{\bf{I}}}&{{{\bf{0}}_{3 \times 3}}}\\
{ - {W_N}\left[ {{{\bf{j}}_1} \times } \right]}&{{W_{N,s}}{\bf{I}}}
\end{array}} \right].
\end{equation}
${W_{N,s}}$ denotes the differentiation of NURBS basis function with respect to the arc-length coordinate, derived as ${W_{N,s}} \equiv {W_{I,\xi }}/{J_c}$ \cite{choi2016isogeometric}. ${(\bullet)_{,\xi }}$ denotes the partial differentiation with respect to the parametric coordinate $\xi $. Then the strain energy bilinear form of Eq.\,(\ref{ig_vib_strn_e_form}) is discretized by
\begin{align} \label{ig_vib_strn_e_form_disc}
&a({{\boldsymbol{\eta }}^h},{{\boldsymbol{\bar \eta }}^h})\nonumber\\
&= {\bf{A}}_{e = 1}^{\text{ne}}\left[ {{{{\bf{\bar d}}}_N}^T\left\{ {\int_{{\Xi _e}} {\left( {{\bf{\Xi }}_{0N}^h{{\bf{\Pi }}_0}{\bf{C}}{{\bf{\Pi }}_0}^T{\bf{\Xi }}{{_{0M}^h}^T}{J_c}} \right)d\xi } } \right\}{{\bf{d}}_M}} \right]\nonumber\\
&\equiv {{\bf{\bar d}}^T}{\bf{Kd}},
\end{align}
where ${{\bf{d}}_N} \equiv {[{{\bf{y}}_N}^T,{{\boldsymbol{\theta }}_N}^T]^T}$ and ${{\bf{\bar d}}_N} \equiv {[{{\bf{\bar y}}_N}^T,{{\boldsymbol{\bar \theta }}_N}^T]^T}$. The repeated indices $N$ and $M$ imply the summations over the range from 1 to $n$, and $n$ denotes the number of control points having local supports in the knot span ${\Xi _e}$. $\text{ne}$ denotes the number of knot spans. $\bf{K}$, $\bf{d}$, and ${\bf{\bar d}}$ are the assembled global stiffness matrix, the global vectors of response coefficients, and virtual response coefficients, respectively. Also, the kinetic energy form of Eq.\,(\ref{ig_vib_kin_e_form}) is discretized by
\begin{align} \label{ig_vib_kin_e_form_disc}
&d({{\boldsymbol{\eta }}^h},{{\boldsymbol{\bar \eta }}^h})\nonumber\\
&= {\bf{A}}_{e = 1}^{\text{ne}}\left[ {{{{\bf{\bar d}}}_N}^T\left\{ {\int_{{\Xi _e}} {{\left[ {\begin{array}{*{20}{c}}
{\rho A{\bf{I}}}&{{{\bf{0}}_{3 \times 3}}}\\
{{{\bf{0}}_{3 \times 3}}}&{{{\bf{I}}_\rho }}
\end{array}} \right]}{W_N}{W_M}{J_c}d\xi } } \right\}{{\bf{d}}_M}} \right]\nonumber\\
&\equiv {{\bf{\bar d}}^T}{\bf{Md}}.
\end{align}
Thus, using Eqs.\,(\ref{ig_vib_strn_e_form_disc}) and (\ref{ig_vib_kin_e_form_disc}), we have the following discretized forms of the generalized eigenvalue problem of Eq.\,(\ref{ig_vib_eval_prob}). 
\begin{equation} \label{ig_vib_eval_prob_disc}
{\bf{Kd}} = \zeta {\bf{Md}},
\end{equation}
where it is assumed that the eigenvectors are $\bf{M}$-orthonormalized, from Eq.\,(\ref{ig_vib_normalize}), as
\begin{equation} \label{ig_vib_normal_condi_disc}
{{\bf{d}}^T}{\bf{Md}} = 1.
\end{equation}
It is noted that a rotation continuity condition at the interface between NURBS curve patches can be automatically satisfied by a typical assembly process, since we employ the rotation vectors as the control coefficients in Eqs.\,(\ref{ig_vib_disp_rot_amp_disc}) and (\ref{ig_vib_disp_rot_amp_vir_disc}).

\subsection{Geometric modeling of embedded structures: review}
\label{modeling_embed_review}
\begin{figure*}[htp]
\centering
\begin{minipage}{0.285\textwidth}
   \subfloat[Initial parent domain]{\label{ig_vib_init_par_domain}
      \includegraphics[width=0.8\textwidth]{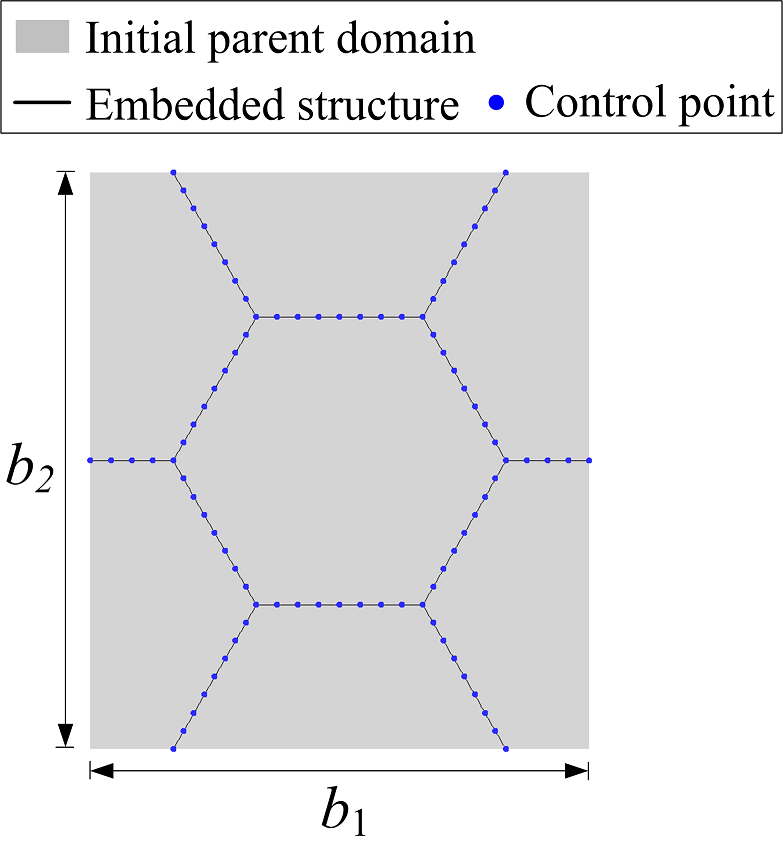}}
\end{minipage}%
\begin{minipage}{0.225\textwidth}
   \subfloat[Modeling by FFD]{\label{ig_vib_model_by_ffd}
      \includegraphics[width=0.8\textwidth]{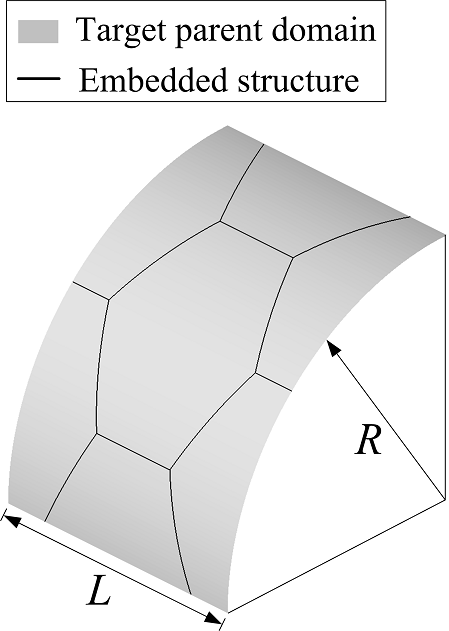}}
\end{minipage}%
\begin{minipage}{0.3\textwidth}
   \subfloat[Curve interpolation]{\label{ig_vib_curve_interp}
      \includegraphics[width=0.8\textwidth]{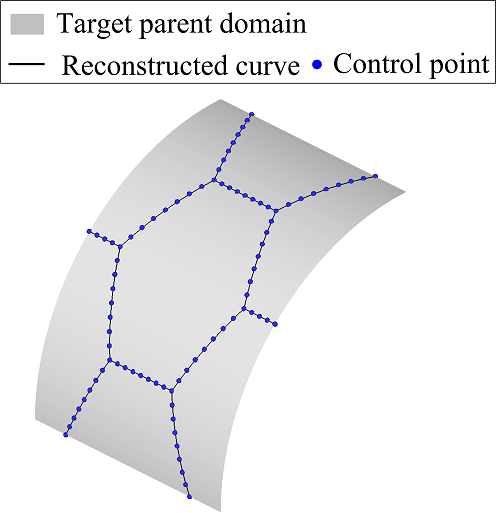}}
\end{minipage}%
   \caption{Geometric modeling of an embedded structure}\label{ig_vib_model_embed_structure}
\end{figure*}

In the previous work of \cite{choi2018constrained}, geometric modeling of constrained beam structure on a curved surface was conducted by combining a free-form deformation (FFD) and a global curve interpolation. Here, we briefly explain the overall procedure. For example, we consider constructing a beam structure on a cylindrical surface, as illustrated in Fig. \ref{ig_vib_model_embed_structure}. We first construct a structure, as shown in Fig. \ref{ig_vib_init_par_domain}, in a rectangular domain with width ${b_1}$ and height ${b_2}$, so-called the ``\textit{initial}'' parent domain ${\tilde \Omega ^0}$. The embedded structure in the initial parent domain is expressed by a NURBS curve of Eq.\,(\ref{ig_vib_nurbs_curve}), as
\begin{equation} \label{ig_vib_embed_curve_init_par}
{{\bf{X}}^0}(\xi ) \equiv \sum\limits_{I = 1}^n {{W_I}(\xi ){\bf{B}}_I^0}  \in {\tilde \Omega ^0}.
\end{equation}
${\bf{B}}_I^0$ and $n$ define a control point position and the number of control points, respectively. We also define a ``\textit{target}'' parent domain ($\tilde \Omega $) like the cylindrical surface having radius $R$ and length $L$ shown in Fig. \ref{ig_vib_model_by_ffd} whose geometry is described by a NURBS surface as
\begin{equation} \label{ig_vib_embed_curve_tar_par}
{\bf{\tilde X}}({\tilde \xi ^1},{\tilde \xi ^2}) = \sum\limits_{J = 1}^m {{{\tilde W}_J}({{\tilde \xi }^1},{{\tilde \xi }^2}){{{\bf{\tilde B}}}_J}},
\end{equation}
where ${\tilde W_J}({\tilde \xi ^1},{\tilde \xi ^2})$ and ${\tilde \xi ^i}\,(i = 1,2)$ are the bivariate NURBS basis function and surface parametric coordinates, respectively, and ${{\bf{\tilde B}}_J}$ defines the control point position, and $m$ denotes the number of control points. An FFD states that the embedded object in an initial parent domain is mapped onto a target parent domain, as shown in Fig. \ref{ig_vib_model_by_ffd} in a way that \cite{choi2018constrained}
\begin{align}
{\bf{X}}(\xi ) &\equiv {\bf{\tilde X}}({{\tilde \xi }^1}(\xi ),{{\tilde \xi }^2}(\xi ))\nonumber\\
 &= \sum\limits_{J = 1}^m {{{\tilde W}_J}({{\tilde \xi }^1}(\xi ),{{\tilde \xi }^2}(\xi )){{{\bf{\tilde B}}}_J}} ,
\label{ig_vib_embed_curve_exact_geom}
\end{align}
where the surface parametric position $({\tilde \xi ^1},{\tilde \xi ^2})$ corresponding to the curve parametric position $\xi $ is determined by a simple algebra as
\begin{equation} \label{ig_vib_embed_curve_param_pos_algebra}
{\tilde \xi ^i} = X_i^0(\xi )/{b_i}, i = 1,2,
\end{equation}
and $X_i^0$ denotes $i$-th component of ${{\bf{X}}^0}$. It is noted that the embedded geometry ${\bf{X}}$ is exactly on the target parent domain; however, in order to determine a corresponding control net, a global curve interpolation needs to be additionally performed, which solves the following system of linear equations \cite{choi2018constrained}.
\begin{equation} \label{ig_vib_embed_interpolate}
{{\bf{N}}_g}{{\bf{B}}_g} = {{\bf{X}}_g},
\end{equation}
where an invertible matrix ${{\bf{N}}_g}$ and the right-hand side vector ${{\bf{X}}_g}$ are respectively constructed by assembling the evaluations of the NURBS basis function ${W_I}(\xi _L^c)$ and embedded curve position ${\bf{X}}(\xi _L^c)$ by Eq. (\ref{ig_vib_embed_curve_exact_geom}) at selected parametric positions $\xi _L^c$ ($L=1,...,n$). From the solution ${{\bf{B}}_g}$ of Eq. (\ref{ig_vib_embed_interpolate}), we extract a set of control point positions $\left\{ {{{\bf{B}}_I}} \right\}$ for each patch in the embedded curves on target parent domain (Fig.\,\ref{ig_vib_curve_interp}), then a reconstructed geometry, that is, a beam neutral axis geometry can be expressed by
\begin{equation} \label{ig_vib_embed_geom_recons}
{{\bf{X}}^ * }(\xi ) = \sum\limits_{I = 1}^n {{W_I}(\xi ){{\bf{B}}_I}}.
\end{equation}
In \cite{choi2018constrained}, it was shown that a deviation of the reconstructed geometry ${{\bf{X}}^ * }(\xi )$ from the target one ${\bf{X}}(\xi )$, so-called a ``\textit{modeling error}'', is able to be reduced by increasing the degrees of freedom (DOFs) of the reconstructed curve through a knot insertion or a degree-elevation process. It was also investigated that the local non-smooth region of target parent domain can be effectively resolved by the multi-patch modeling of beam structures. For more detailed discussions regarding the way to reduce the modeling error, one can refer to the reference \cite{choi2018constrained}.

The configuration design DOFs of the reconstructed curve, shown in Fig. \ref{ig_vib_curve_interp}, can be divided into two parts: first, the geometry of embedded lattice structure in initial parent domain (e.g., Fig.\,\ref{ig_vib_init_par_domain}), and second, the geometry of target parent domain, for example, the cylindrical surface shown in Fig. \ref{ig_vib_model_by_ffd} may change its radius ($R$) and length ($L$). In the previous work of \cite{choi2018constrained}, only the configuration design of embedded structures in the initial parent domain was altered, that is, a structure was constrained on a given specific surface. \textit{However, in this paper, we extend the previous work to consider a design dependence of target parent domain, so that we have additional design DOFs.} In the following section, we explain the design velocity computation considering the design dependence of target parent domain.

\section{Configuration DSA of eigenvalues}
\label{ig_vib_config_dsa}
\subsection{Computation of design velocity field}
We present a continuum-based analytical configuration DSA of eigenvalues. In this paper, it is assumed that a beam cross-section has a circular shape; thus, an initial cross-section orientation is determined by only a neutral axis configuration design. The neutral axis position is perturbed by a given design velocity field ${\bf{V}}(s)$ as \cite{choi2016isogeometric}
\begin{equation} \label{ig_vib_neut_axis_ptb}
{{\boldsymbol{\varphi }}_{0\tau }}({s_\tau }) \equiv {{\boldsymbol{\varphi }}_0}(s) + \tau {\bf{V}}(s),
\end{equation}
where $\tau  \ge 0$ represents a design perturbation amount which is a time-like parameter that controls a design variation, and the subscript $\tau $ represents a quantity evaluated at a perturbed design here and hereafter. Taking the material derivative of Eq.\,(\ref{ig_vib_neut_axis_ptb}) gives
\begin{equation} \label{ig_vib_neut_axis_pos_mat_deriv}
{{\boldsymbol{\dot \varphi }}_0}(s) \equiv \frac{d}{{d\tau }}{\left. {{{\boldsymbol{\varphi }}_{0\tau }}({s_\tau })} \right|_{\tau  = 0}} = {\bf{V}}(s),
\end{equation}
and the upper dot $(\dot{\bullet})$ denotes the material derivative, which is sometimes substituted by a superscript dot ${(\bullet)^{\sbt}}$.

We separate a configuration design variables into two sets. \textit{First}, configuration changes of an embedded structure in the initial parent domain are described by a set of design variables, and each variable has a given design velocity field, expressed by \cite{choi2018constrained}
\begin{align} \label{ig_vib_des_vel_init_par}
{{\bf{V}}^0}(\xi ) &\equiv \frac{d}{{d\tau }}{\left. {{\bf{X}}_\tau ^0} \right|_{\tau  = 0}}\nonumber\\
&= \sum\limits_{I = 1}^n {{W_I}(\xi ){\bf{\dot B}}_I^0}\nonumber\\
&= \sum\limits_{I = 1}^n {{W_I}(\xi ){\bf{V}}_I^0},
\end{align}
where ${\bf{V}}_I^0 \equiv {\bf{\dot B}}_I^0$ represents a given design velocity coefficient. For a given design velocity field of Eq.\,(\ref{ig_vib_des_vel_init_par}), the material derivative of surface parametric coordinates of Eq.\,(\ref{ig_vib_embed_curve_param_pos_algebra}) is obtained by \cite{choi2018constrained}
\begin{equation} \label{ig_vib_mat_deriv_param_coord_eobj}
{{{\dot{\tilde \xi}^i}}} = V_i^0(\xi )/{b_i},\,\,\,\,i=1,2,
\end{equation}
where $V_i^0$ denotes the $i$-th component of ${{\bf{V}}^0}$. \textit{Second}, the target parent domain has a set of design variables, and each variable has a given design velocity field, expressed as
\begin{align} \label{ig_vib_des_vel_tar_par}
{\bf{\tilde V}}({\tilde \xi ^1},{\tilde \xi ^2}) &\equiv \frac{d}{{d\tau }}{\left. {{{{\bf{\tilde X}}}_\tau }} \right|_{\tau  = 0}}\nonumber\\
&= \sum\limits_{J = 1}^m {{{\tilde W}_J}({{\tilde \xi }^1},{{\tilde \xi }^2}){{{\bf{\dot {\tilde B}}}}_J}} \nonumber\\ 
&= \sum\limits_{J = 1}^m {{{\tilde W}_J}({{\tilde \xi }^1},{{\tilde \xi }^2}){{{\bf{\tilde V}}}_J}},
\end{align}
where ${{\bf{\tilde V}}_J}$ denotes a given design velocity coefficient. Taking the material derivative of the surface position of Eq.\,(\ref{ig_vib_embed_curve_exact_geom}) yields a design velocity vector on the target parent domain as
\begin{equation} \label{ig_vib_des_vel_on_tar_par_domain}
{\bf{\dot X}}(\xi ) = \sum\limits_{J = 1}^m {\left( {{{{\tilde W}_{J,{{\tilde \xi }^i}}}{{\dot {\tilde {\xi}}^i}}}{{{\bf{\tilde B}}}_J} + {{\tilde W}_J}{{{\bf{\tilde V}}}_J}} \right)}\equiv {\bf{V}}(\xi)
\end{equation}
where the repeated index $i$ implies summation over $1$ and $2$. ${(\bullet)_{,{{\tilde \xi }^i}}}$ denotes the differentiation with respect to the surface parametric coordinates ${\tilde \xi ^i}$ ($i=1,2$). Similar to the geometric modeling procedure in section \ref{modeling_embed_review}, we interpolate the design velocity vectors of Eq. (\ref{ig_vib_des_vel_on_tar_par_domain}) to have the corresponding control coefficients by solving the following system of linear equations
\begin{equation} \label{ig_vib_des_vel_interpolate}
{{\bf{N}}_g}{{\bf{V}}_g} = {{\bf{\dot X}}_g},
\end{equation}
where the same system matrix ${{\bf{N}}_g}$ with that of Eq. (\ref{ig_vib_embed_interpolate}) is used, and the right-hand side is constructed by an assembly of evaluations ${\bf{\dot X}}(\xi _L^c)$ of Eq. (\ref{ig_vib_des_vel_on_tar_par_domain}) at selected $n$ discrete points $\xi  = \xi _L^c$ ($L=1,...,n$). From the solution ${{\bf{V}}_g}$ of Eq. (\ref{ig_vib_des_vel_interpolate}), we extract a set of design velocity coefficients $\left\{ {{{\bf{V}}_I}} \right\}$ for each patch in the embedded curves on the target parent domain, then the design velocity field is finally expressed as
\begin{equation} \label{ig_vib_des_vel_recon}
{\bf{V}}(s(\xi )) \equiv \frac{d}{{d\tau }}{\left. {{\bf{X}}_\tau ^ * } \right|_{\tau  = 0}} = \sum\limits_{I = 1}^n {{W_I}(\xi ){{\bf{V}}_I}},
\end{equation}
where ${{\bf{V}}_I}$ denotes a design velocity coefficient at each control point. 
\subsection{Material derivative of initial triads in embedded structures}
For a $C^1$-continuous spatial curve, a unit tangent vector ${{\bf{j}}_1}(\xi )$ can be uniquely determined along the curve. However, the other two base vectors ${{\bf{j}}_2}(\xi )$ and ${{\bf{j}}_3}(\xi )$ of the orthonormal basis are not uniquely determined due to the rotational DOF around the curve. In order to explicitly parameterize local orthonormal frames, the SR method was employed in \cite{meier2014objective}, as
\begin{align} \label{ig_vib_base_vec_sr_method}
{{\bf{j}}_k}(\xi ) & = {\bf{j}}_k^{\text{ref}} - \frac{{{\bf{j}}{{_k^{\text{ref}}}^T}{{\bf{j}}_1}(\xi )}}{{1 + {\bf{j}}{{_1^{\text{ref}}}^T}{{\bf{j}}_1}(\xi )}}\left\{ {{{\bf{j}}_1}(\xi ) + {\bf{j}}_1^{\text{ref}}} \right\} \nonumber\\
& \equiv {{\bf{s}}_k}\left( {{{\bf{j}}_1}(\xi ),{\bf{j}}_{123}^{\text{ref}}} \right)\,\,,\,\,k=2,3,
\end{align}
where ${\bf{j}}_{123}^{\text{ref}} \equiv \left\{ {{\bf{j}}_1^{\text{ref}},{\bf{j}}_2^{\text{ref}},{\bf{j}}_3^{\text{ref}}} \right\}$ denotes the reference basis, and ${{\bf{s}}_k}(\bullet,*)$ denotes the SR operation which is interpreted as rotating a given reference base vectors (*) in a way that the first reference base vector (${\bf{j}}_1^{\text{ref}}$) is rotated to a given vector ($\bullet$) with a minimal rotation angle. The SR method was also employed in \cite{choi2019isogeometric} within the IGA framework, and a surface convected basis was shown to be effectively utilized as the reference basis for curves embedded in a smooth surface. In this section, we explain an extension of the previous work to derive material derivatives of initial orthonormal base vectors, considering the design dependence of target parent domain.

If a curve and its perturbed design are $C^1$-continuous, the material derivative ${\bf{j}}_1^{\sbt}(\xi )$ can be obtained by taking the material derivative of ${{\bf{j}}_1} \equiv {{\boldsymbol{\varphi }}_{0,s}}$ as \cite{choi2019isogeometric}
\begin{equation} \label{ig_vib_mat_deriv_j1}
{\bf{j}}_1^{\sbt}(\xi ) = {{\bf{V}}_{,s}} - {{\bf{j}}_1}{\nabla _s} \cdot {\bf{V}},
\end{equation}
where ${\nabla _s} \cdot ( \bullet ) \equiv {( \bullet )_{,s}} \cdot {{\bf{j}}_1}$. The other two base vectors ${{\bf{j}}_2}(\xi )$ and ${{\bf{j}}_3}(\xi )$ of the orthonormal basis can be determined by \cite{choi2019isogeometric}
\begin{equation} \label{ig_vib_sr_method_pw}
{\bf{j}}_k^{}(\xi ) = {{\bf{s}}_k}\left( {{{\bf{j}}_1}(\xi ),{\bf{j}}_{123}^{\text{ref}}(\xi )} \right),\,\,k=2,3,
\end{equation}
where ${\bf{j}}_{123}^{\text{ref}}(\xi ) \equiv \left\{ {{\bf{j}}_1^{\text{ref}}(\xi ),{\bf{j}}_2^{\text{ref}}(\xi ),{\bf{j}}_3^{\text{ref}}(\xi )} \right\}$ represents the reference orthonormal basis, defined by a surface convected basis in the following. The SR reference tangent vector, which is an exact tangent to the target parent domain along the embedded curve, is defined as \cite{choi2019isogeometric}
\begin{equation} \label{ig_vib_j1_ref_pw}
{\bf{j}}_1^{\text{ref}}(\xi ) \equiv \frac{{\partial {\bf{\tilde X}}/\partial \xi }}{{\left\| {\partial {\bf{\tilde X}}/\partial \xi } \right\|}},
\end{equation}
where
\begin{align} \label{ig_vib_partial_deriv_tar_par}
\frac{\partial }{{\partial \xi }}{\bf{\tilde X}}&\equiv {{\bf{\tilde X}}_{,{{\tilde \xi }^1}}}\frac{{\partial {{\tilde \xi }^1}}}{{\partial \xi }} + {{\bf{\tilde X}}_{,{{\tilde \xi }^2}}}\frac{{\partial {{\tilde \xi }^2}}}{{\partial \xi }}\nonumber\\
&= \frac{1}{{{b_1}}}X_{1,\xi }^0{{\bf{\tilde X}}_{,{{\tilde \xi }^1}}} + \frac{1}{{{b_2}}}X_{2,\xi }^0{{\bf{\tilde X}}_{,{{\tilde \xi }^2}}}.
\end{align}
The SR reference unit binormal and normal vectors are respectively determined by \cite{choi2019isogeometric}
\begin{equation} \label{ig_vib_j3_ref_pw}
{\bf{j}}_3^{\text{ref}}(\xi ) \equiv \frac{{{{{\bf{\tilde X}}}_{,{{\tilde \xi }^1}}} \times {{{\bf{\tilde X}}}_{,{{\tilde \xi }^2}}}}}{{\left\| {{{{\bf{\tilde X}}}_{,{{\tilde \xi }^1}}} \times {{{\bf{\tilde X}}}_{,{{\tilde \xi }^2}}}} \right\|}},
\end{equation}
and
\begin{equation} \label{ig_vib_j2_ref}
{\bf{j}}_2^{\text{ref}}(\xi ) \equiv {\bf{j}}_3^{\text{ref}}(\xi ) \times {\bf{j}}_1^{\text{ref}}(\xi ).
\end{equation}
Taking the material derivative to Eq. (\ref{ig_vib_j1_ref_pw}) gives
\begin{equation} \label{ig_vib_j1_ref_dot}
{\bf{j}}_1^{\text{ref}{\sbt}}(\xi ) = \frac{1}{{\left\| {\partial {\bf{\tilde X}}/\partial \xi } \right\|}}{{\bf{P}}_{{\bf{j}}_1^{\text{ref}}}}\frac{{\partial {\bf{\dot {\tilde X}}}}}{{\partial \xi }},
\end{equation}
where ${{\bf{P}}_{(\bullet)}} \equiv {\bf{I}} - (\bullet) \otimes (\bullet)$, and $\bf{I}$ denotes the identity matrix, and $\partial {\bf{\dot {\tilde X}}}/\partial \xi$ can be explicitly expressed in terms of the given design velocity fields of Eqs.\,(\ref{ig_vib_des_vel_init_par}) and (\ref{ig_vib_des_vel_tar_par}) from Eq.\,(\ref{ig_vib_partial_deriv_tar_par}) as 
\begin{align}\label{deriv_x_td_dot}
\frac{\partial }{{\partial \xi }}{\bf{\dot {\tilde X}}}({\tilde \xi ^1},{\tilde \xi ^2})&=\frac{1}{{{b_1}}}\left\{ {V_{1,\xi }^0{{{\bf{\tilde X}}}_{,{{\tilde \xi }^1}}} + X_{1,\xi }^0{{\left( {{{{\bf{\tilde X}}}_{,{{\tilde \xi }^1}}}} \right)}^{\sbt}}} \right\}\nonumber\\
&+ \frac{1}{{{b_2}}}\left\{ {V_{2,\xi }^0{{{\bf{\tilde X}}}_{,{{\tilde \xi }^2}}} + X_{2,\xi }^0{{\left( {{{{\bf{\tilde X}}}_{,{{\tilde \xi }^2}}}} \right)}^{\sbt}}} \right\}.
\end{align}
We also take the material derivative of Eq.\,(\ref{ig_vib_j3_ref_pw}) to have
\begin{align} \label{ig_vib_mat_deriv_j3_ref}
{\bf{j}}_3^{{\text{ref}} \sbt }&= \frac{1}{{\left\| {{{{\bf{\tilde X}}}_{,{{\tilde \xi }^1}}} \times{{{\bf{\tilde X}}}_{,{{\tilde \xi }^2}}}} \right\|}}\times\nonumber\\
&{{\bf{P}}_{{\bf{j}}_3^{\text{ref}}}}\left\{ {{{\left( {{{{\bf{\tilde X}}}_{,{{\tilde \xi }^1}}}} \right)}^{\sbt}} 
\times {{{\bf{\tilde X}}}_{,{{\tilde \xi }^2}}} + {{{\bf{\tilde X}}}_{,{{\tilde \xi }^1}}} \times {{\left( {{{{\bf{\tilde X}}}_{,{{\tilde \xi }^2}}}} \right)}^{\sbt}}} \right\}.
\end{align}
In Eqs.\,(\ref{deriv_x_td_dot}) and (\ref{ig_vib_mat_deriv_j3_ref}), ${\left( {{{{\bf{\tilde X}}}_{,{{\tilde \xi }^i}}}} \right)^{\sbt}}$ ($i=1,2$) can be calculated by
\begin{equation}
{\left( {{{{\bf{\tilde X}}}_{,{{\tilde \xi }^i}}}} \right)^{\sbt}} = {{\bf{\tilde V}}_{,{{\tilde \xi }^i}}} + \frac{{V_1^0}}{{{b_1}}}{{\bf{\tilde X}}_{,{{\tilde \xi }^i}{{\tilde \xi }^1}}} + \frac{{V_2^0}}{{{b_2}}}{{\bf{\tilde X}}_{,{{\tilde \xi }^i}{{\tilde \xi }^2}}}.
\end{equation}
The material derivative of the second reference vector is obtained from Eq.\,(\ref{ig_vib_j2_ref}), as \cite{choi2019isogeometric}
\begin{equation} \label{ig_vib_mat_deriv_j2_ref}
\begin{array}{c}
{\bf{j}}_2^{{\text{ref}} \sbt }(\xi ) \equiv {\bf{j}}_3^{{\text{ref}} \sbt }(\xi ) \times {\bf{j}}_1^{{\text{ref}}}(\xi ) + {\bf{j}}_3^{{\text{ref}}}(\xi ) \times {\bf{j}}_1^{{\text{ref}} \sbt }(\xi ).
\end{array}
\end{equation}
By taking the material derivative of the SR operation expression of Eq.\,(\ref{ig_vib_sr_method_pw}), the following was obtained \cite{choi2019isogeometric}. 
\begin{align}\label{ig_vib_mat_deriv_j23}
{\bf{j}}_k^{\sbt} &= {\bf{j}}_k^{{\text{ref}} {\sbt} }- \frac{1}{{1 + {\bf{j}}_1^{\text{ref}}{^T}{{\bf{j}}_1}}}\left[ {\begin{array}{*{20}{l}}
{\left( {{\bf{j}}_k^{{\text{ref}}{\sbt} }{^T}{{\bf{j}}_1} + {\bf{j}}_k^{\text{ref}}{^T}{\bf{j}}_1^ {\sbt}} \right)\left( {{{\bf{j}}_1} + {\bf{j}}_1^{{\text{ref}}}} \right)}\\
{ - \left( {{\bf{j}}_1^{{\text{ref}}{\sbt}}{^T}{{\bf{j}}_1} + {\bf{j}}_1^{\text{ref}}{^T}{\bf{j}}_1^{\sbt}} \right)\left( {{\bf{j}}_k^{\text{ref}} - {{\bf{j}}_k}} \right)}\\
{ + {\bf{j}}_k^{\text{ref}}{^T}{{\bf{j}}_1}\left( {{\bf{j}}_1^{\sbt} + {\bf{j}}_1^{{\text{ref}}{\sbt} }} \right)}
\end{array}} \right]\nonumber\\
&= {\bf{s}}_k^{\sbt} \left( {{{\bf{j}}_1},{\bf{j}}_{123}^{\text{ref}};{\bf{j}}_1^{\sbt},{\bf{j}}_{123}^{{\text{ref}}{\sbt}}} \right)\,\,,k = 2,3. 
\end{align}
where ${\bf{j}}_{123}^{{\text{ref}}{\sbt}}\equiv \left\{ {{\bf{j}}_1^{{\text{ref}}{\sbt}},{\bf{j}}_2^{{\text{ref}}{\sbt}},{\bf{j}}_3^{{\text{ref}}{\sbt}}} \right\}$ is obtained by Eqs.\,(\ref{ig_vib_j1_ref_dot}), (\ref{ig_vib_mat_deriv_j3_ref}), and (\ref{ig_vib_mat_deriv_j2_ref}). Then we can calculate ${{\bf{\dot \Lambda }}_0} = \left[ {{\bf{j}}_1^{\sbt},{\bf{j}}_2^{\sbt},{\bf{j}}_3^{\sbt}} \right]$ using Eqs.\,(\ref{ig_vib_mat_deriv_j1}) and (\ref{ig_vib_mat_deriv_j23}). The material derivatives ${\bf{j}}_I^{{\text{ref}}{\sbt}}$ ($I = 1,2,3$) reflect a configuration design variation of target parent domain by the terms regarding a given design velocity field of Eq.\,(\ref{ig_vib_des_vel_tar_par}), in contrast to the corresponding expressions in \cite{choi2019isogeometric} where the designs of target parent domains were not varied.
\subsection{Material derivatives of strain measures}
For future use, we recall the expression of the material derivative of the infinitesimal arclength in \cite{choi2016isogeometric}, as
\begin{equation} \label{ig_vib_mat_deriv_ds}
{(ds)^ {\sbt}} = {\nabla _s} \cdot {\bf{V}}ds.
\end{equation}
We also recall the material derivatives of the linearized strain measures from \cite{choi2018constrained}.
\begin{align} \label{ig_vib_mat_deriv_strn_gamma}
{\left\{ {{\bf{\Gamma }}({\boldsymbol{\eta }})} \right\}^{\sbt}} & = {\bf{\Lambda }}_0^T({{\bf{\dot z}}_{,s}} - {\boldsymbol{\dot \theta }} \times {{\bf{j}}_1}) \nonumber\\
& +  {{\bf{\Lambda }}_0^T( - {{\bf{z}}_{,s}}{\nabla _s} \cdot {\bf{V}} + {\bf{j}}_1^{\sbt}  \times {\boldsymbol{\theta }}) + {{{\bf{\dot \Lambda }}}_0}^T{{\bf{\Lambda }}_0}{\bf{\Gamma }}}  \nonumber\\ 
& \equiv {\bf{\Gamma }}({\boldsymbol{\dot \eta }}) + {{\bf{\Gamma '}}_V}({\boldsymbol{\eta }}),
\end{align}
and
\begin{align} \label{ig_vib_mat_deriv_strn_omega}
{\left\{ {{\bf{\Omega }}({\bf{\eta }})} \right\}^{\sbt}} & = {{\bf{\Lambda }}_0}^T{{\boldsymbol{\dot \theta }}_{,s}} + \left\{ {{{\bf{\Lambda }}_0}^T( - {{\boldsymbol{\theta }}_{,s}}{\nabla _s} \cdot {\bf{V}}) + {{{\bf{\dot \Lambda }}}_0}^T{{\bf{\Lambda }}_0}{\bf{\Omega }}} \right\}\nonumber\\
& \equiv {\bf{\Omega }}({\boldsymbol{\dot \eta }}) + {{\bf{\Omega '}}_V}({\boldsymbol{\eta }}),
\end{align}
where ${\boldsymbol{\dot \eta }} \equiv ({\bf{\dot z}}(s),{\boldsymbol{\dot \theta }}(s))$. Then, taking the material derivative of Eq. (\ref{ig_vib_strn_e_form}), using Eqs. (\ref{ig_vib_mat_deriv_ds})-(\ref{ig_vib_mat_deriv_strn_omega}), yields \cite{choi2018constrained} 
\begin{align}\label{ig_vib_mat_deriv_strn_e_form}
&{[a({\boldsymbol{\eta }},{\boldsymbol{\bar \eta }})]^{\sbt}} = \int_\Omega  {\left\{ {{\bf{\Gamma }}{{({\boldsymbol{\dot {\bar \eta}}})}^T}{{\bf{C}}_F}{\bf{\Gamma }}({\boldsymbol{\eta }}) + {\boldsymbol{\Omega }}{{({\boldsymbol{\dot {\bar \eta}}})}^T}{{\bf{C}}_M}{\bf{\Omega }}({\boldsymbol{\eta }})} \right\}ds} \nonumber\\
&+ \int_\Omega  {\left\{ {{\bf{\Gamma }}{{({\boldsymbol{\bar \eta }})}^T}{{\bf{C}}_F}{\bf{\Gamma }}({\boldsymbol{\dot \eta }}) + {\bf{\Omega }}{{({\boldsymbol{\bar \eta }})}^T}{{\bf{C}}_M}{\bf{\Omega }}({\boldsymbol{\dot \eta }})} \right\}ds}\nonumber\\
&+ \int_\Omega  {\left[ \begin{array}{l}
{{{\bf{\Gamma '}}}_V}{({\boldsymbol{\bar \eta }})^T}{{\bf{C}}_F}{\bf{\Gamma }}({\boldsymbol{\eta }})\\
 + {{{\bf{\Omega '}}}_V}{({\boldsymbol{\bar \eta }})^T}{{\bf{C}}_M}{\bf{\Omega }}({\boldsymbol{\eta }})\\
 + {\bf{\Gamma }}{({\boldsymbol{\bar \eta }})^T}{{\bf{C}}_F}{{{\bf{\Gamma '}}}_V}({\boldsymbol{\eta }})\\
 + {\bf{\Omega }}{({\boldsymbol{\bar \eta }})^T}{{\bf{C}}_M}{{{\bf{\Omega '}}}_V}({\boldsymbol{\eta }})\\
 + \left\{ \begin{array}{l}
{\bf{\Gamma }}{({\boldsymbol{\bar \eta }})^T}{{\bf{C}}_F}{\bf{\Gamma }}({\boldsymbol{\eta }})\\
 + {\bf{\Omega }}{({\boldsymbol{\bar \eta }})^T}{{\bf{C}}_M}{\bf{\Omega }}({\boldsymbol{\eta }})
\end{array} \right\}{\nabla _s} \cdot {\bf{V}}
\end{array} \right]ds}\nonumber\\
&\equiv a({\boldsymbol{\eta }},{\boldsymbol{\dot {\bar \eta}}}) + a({\boldsymbol{\dot \eta }},{\boldsymbol{\bar \eta }}) + {a'_V}({\boldsymbol{\eta }},{\boldsymbol{\bar \eta }}),
\end{align}
where ${\boldsymbol{\dot {\bar \eta}}} \equiv ({\bf{\dot {\bar z}}}(s),{\boldsymbol{\dot {\bar \theta}}}(s))$. Also, taking the material derivative of Eq. (\ref{ig_vib_kin_e_form}) leads to 
\begin{align}\label{ig_vib_mat_deriv_kin_e_form}
{\left[ {d({\boldsymbol{\eta }},{\boldsymbol{\bar \eta }})} \right]^{\sbt} }  &= \int_\Omega  {\left( {{{{\bf{\dot {\bar z}}}}^T}\rho A{\bf{z}} + {{{\boldsymbol{\dot {\bar \theta }}^T}}}{{\bf{I}}_\rho }{\boldsymbol{\theta }}} \right)ds}  \nonumber\\
& + \int_\Omega  {\left( {{{{\bf{\bar z}}}^T}\rho A{\bf{\dot z}} + {{\boldsymbol{\bar \theta }}^T}{{\bf{I}}_\rho }{\boldsymbol{\dot \theta }}} \right)ds}   \nonumber\\
&  + \int_\Omega  {\left\{ {{{{\boldsymbol{\bar \theta }}}^T}{{{\bf{\dot I}}}_\rho }{\boldsymbol{\theta }} + \left( {{{{\bf{\bar z}}}^T}\rho A{\bf{z}} + {{{\boldsymbol{\bar \theta }}}^T}{{\bf{I}}_\rho }{\boldsymbol{\theta }}} \right){\nabla _s} \cdot {\bf{V}}} \right\}ds}  \nonumber\\
& \equiv d({\boldsymbol{\eta }},{\boldsymbol{\dot {\bar \eta} }}) + d({\boldsymbol{\dot \eta }},{\boldsymbol{\bar \eta }}) + {d'_V}({\boldsymbol{\eta }},{\boldsymbol{\bar \eta }}),
\end{align}
where ${{\bf{I}}_\rho }$ of Eq. (\ref{ig_vib_inert_tensor}) has a configuration design dependence through ${{\bf{\Lambda }}_0}$, so that ${{\bf{\dot I}}_\rho }$ is obtained by
\begin{align} \label{ig_vib_mat_deriv_kin_e_form}
{{\bf{\dot I}}_\rho } &\equiv {{\bf{\dot \Lambda }}_0}diag{[\rho {I_p},\rho {I_2},\rho {I_3}]_{\left\{ {{{\bf{e}}_I}} \right\}}}{{\bf{\Lambda }}_0}^T\nonumber\\
&+ {{\bf{\Lambda }}_0}diag{[\rho {I_p},\rho {I_2},\rho {I_3}]_{\left\{ {{{\bf{e}}_I}} \right\}}}{{\bf{\dot \Lambda }}_0}^T.
\end{align}
It is noted that the explicit design dependence terms ${a'_V}({\boldsymbol{\eta }},{\boldsymbol{\bar \eta }})$ and ${d'_V}({\boldsymbol{\eta }},{\boldsymbol{\bar \eta }})$ are linear with respect to eigenfunctions and the design velocity field of Eq.\,(\ref{ig_vib_des_vel_recon}).
\subsection{DSA of simple eigenvalues}
In case of a simple eigenvalue, the eigenvalue and the corresponding eigenfunction are Fr{\'e}chet differentiable \cite{haug1986design}. Evaluating Eq.\,(\ref{ig_vib_eval_prob}) at ${\boldsymbol{\bar \eta }} = {\boldsymbol{\dot \eta }}$ since they belong to the same function space $\bar Z$, and using the symmetries of strain and kinetic energy bilinear forms, we obtain the following identity
\begin{equation} \label{ig_vib_eval_prob_eval_eta_dot}
a({\boldsymbol{\dot \eta }},{\boldsymbol{\eta }}) = \zeta d({\boldsymbol{\dot \eta }},{\boldsymbol{\eta }}).
\end{equation}
Taking the material derivative of both sides of Eq.\,(\ref{ig_vib_eval_prob}) and rearranging terms gives the following.
\begin{align} \label{ig_vib_dsa_simple_eval_mat_deriv_var_eq}
a({\boldsymbol{\dot \eta }},{\boldsymbol{\bar \eta }}) + {a'_V}({\boldsymbol{\eta }},{\boldsymbol{\bar \eta }}) &= \zeta \left\{ {d({\boldsymbol{\dot \eta }},{\boldsymbol{\bar \eta }}) + {{d'}_V}({\boldsymbol{\eta }},{\boldsymbol{\bar \eta }})} \right\}\nonumber\\
&+ \dot \zeta d({\boldsymbol{\eta }},{\boldsymbol{\bar \eta }}),\,\,\forall {\boldsymbol{\bar \eta }} \in \bar Z.
\end{align}
Evaluating Eq.\,(\ref{ig_vib_dsa_simple_eval_mat_deriv_var_eq}) at ${\boldsymbol{\bar \eta }} = {\boldsymbol{\eta }}$ since they belong to the same function space $\bar Z$, and using the identity of Eq.\,(\ref{ig_vib_eval_prob_eval_eta_dot}) and the normalization condition of Eq.\,(\ref{ig_vib_normalize}), we have the following configuration design sensitivity expression for a simple eigenvalue \cite{ha2006design}.
\begin{equation} \label{ig_vib_dsa_simple_eval_mat_deriv}
\dot \zeta  = {a'_V}({\boldsymbol{\eta }},{\boldsymbol{\eta }}) - \zeta {d'_V}({\boldsymbol{\eta }},{\boldsymbol{\eta }}),
\end{equation}
which is linear in the given design velocity field. 
\subsection{DSA of repeated eigenvalues}
Consider that the generalized eigenvalue problem of Eq. (\ref{ig_vib_eval_prob}) yields $S$-fold repeated eigenvalue
\begin{equation} \label{ig_vib_dsa_mult_eval_S_evals}
\tilde \zeta  = {\zeta ^i},\,\,i = 1,...,S,
\end{equation}
which is associated with its $S$ linearly independent eigenfunctions ${{\boldsymbol{\eta }}^i} \equiv ({{\bf{z}}^i},{{\boldsymbol{\theta }}^i})$, and any linear combinations of those eigenfunctions will also satisfy the Eq.\,(\ref{ig_vib_eval_prob}). It was shown that the repeated eigenvalue is not Fr{\'e}chet differentiable, but only directionally differentiable \cite{haug1986design}. At a perturbed design, the repeated eigenvalue may branch into $S$ eigenvalues as
\begin{equation} \label{ig_vib_dsa_mult_eval_ptb}
\zeta _\tau ^i = \tilde \zeta  + \tau {\mu ^i} + o(\tau ),\,\,i = 1,...,S,
\end{equation}
where ${\mu ^i}$ denote directional derivatives of the repeated eigenvalue, and the term $o(\tau )$ represents a quantity which approaches zero faster than $\tau $, i.e., ${\lim _{\tau  \to 0}}o(\tau )/\tau  = 0$ \cite{haug1986design}. The directional derivatives ${\mu ^i}$ are derived as eigenvalues of a $S \times S$ matrix whose $(i,j)$ component is expressed by \cite{seyranian1994multiple,haug1986design} 
\begin{equation} \label{ig_vib_dsa_mult_dir_deriv}
{H_{ij}} = {a'_V}({{\boldsymbol{\eta }}^i},{{\boldsymbol{\eta }}^j}) - \tilde \zeta {d'_V}({{\boldsymbol{\eta }}^i},{{\boldsymbol{\eta }}^j}),\,\,i,j=1,...,S.
\end{equation}
It is noted that the matrix components of Eq.\,(\ref{ig_vib_dsa_mult_dir_deriv}) are linear in the design velocity field; however, the directional derivatives, i.e., the eigenvalues may not be linear in the design velocity field. If the eigenvalue is simple ($S=1$), the directional derivative reduces to the sensitivity expression of Eq.\,(\ref{ig_vib_dsa_simple_eval_mat_deriv}). 
\section{Numerical examples}
\label{ig_vib_num_ex}
\subsection{Clamped rib structure}
\begin{figure*}[htp]
\centering
\begin{minipage}{0.3\textwidth}
   \subfloat[Boundary condition]{\label{ig_vib_model_des}
      \includegraphics[width=\textwidth]{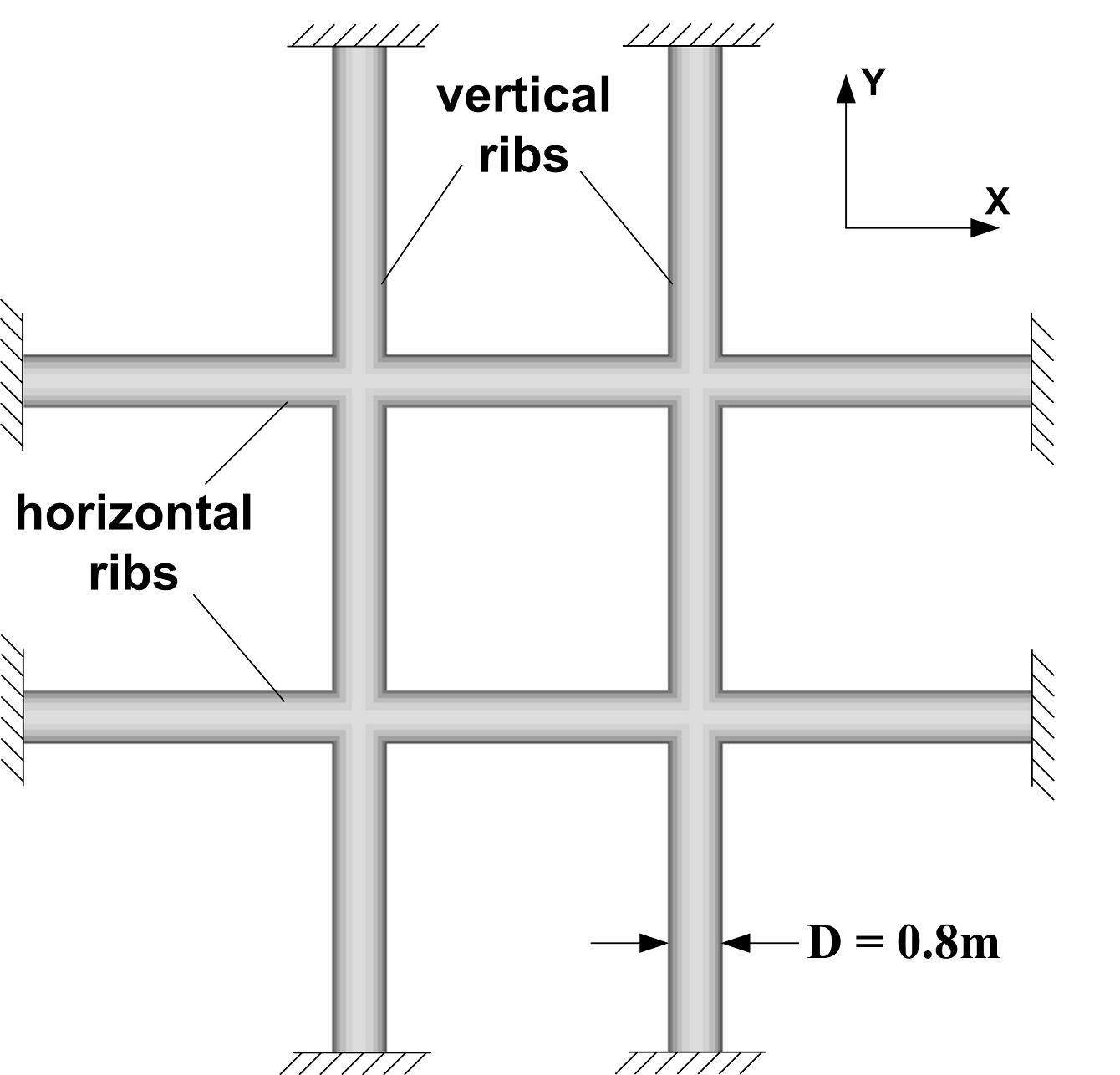}}
\end{minipage}%
\quad\quad
\begin{minipage}{0.3\textwidth}
   \subfloat[Design variables]{\label{ig_vib_des_var}
      \includegraphics[width=\textwidth]{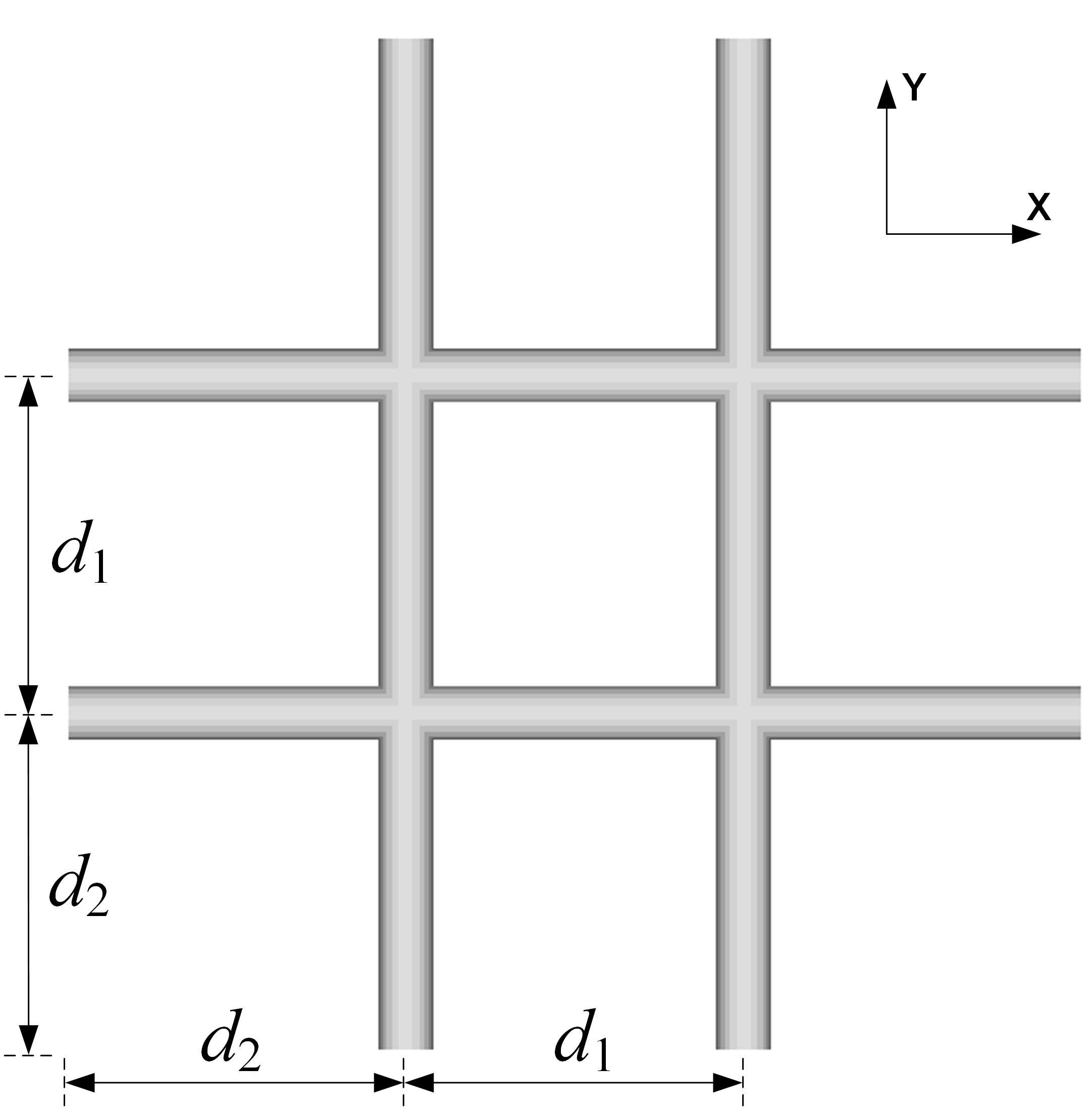}}
\end{minipage}%
   \caption{Model description}\label{ig_vib_model_cant_rib}
\end{figure*}

We consider a clamped rib structural model composed of two horizontal and two vertical ribs, shown in Fig. \ref{ig_vib_model_des}, and the cross-section has a circular shape with diameter $D = 0.8m$, and the material properties are the Young's modulus $E = 200\rm{GPa}$, the Poisson's ratio $\nu  = 0.29$, and the mass density $\rho  = 7,850kg/{m^3}$.

\begin{figure*}[htp]
\centering
\begin{minipage}{0.275\textwidth}
   \subfloat[Mode {\#}1 (${f_1} = 16.68{\rm{Hz}}$)]{\label{ig_vib_cant_rib_mode1}
      \includegraphics[width=0.8\textwidth]{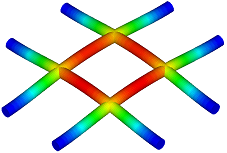}}
\end{minipage}%
\quad\quad
\begin{minipage}{0.275\textwidth}
   \subfloat[Mode {\#}2 (${f_2} = 30.97{\rm{Hz}}$)]{\label{ig_vib_cant_rib_mode2}
      \includegraphics[width=0.8\textwidth]{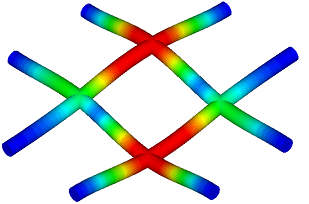}}
\end{minipage}%
\quad\quad
\begin{minipage}{0.345\textwidth}
   \subfloat[Mode {\#}3 (${f_3} = 30.97{\rm{Hz}}$)]{\label{ig_vib_cant_rib_mode3}
      \includegraphics[width=0.8\textwidth]{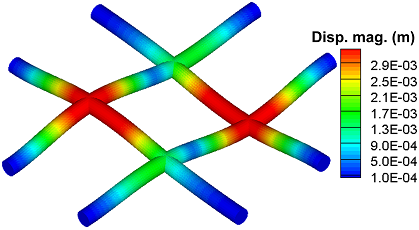}}
\end{minipage}%
   \caption{Mode shapes (scaling factor 100 used for deformed configurations)}\label{ig_vib_cant_rib_mode_shapes}
\end{figure*}

Two design variables ${d_1}$ and ${d_2}$ are considered, as illustrated in Fig. \ref{ig_vib_des_var}; the first one simultaneously changes the horizontal and vertical space between ribs, which maintains the symmetrical design. The second one changes the positions of the lowermost horizontal rib and the leftmost vertical ribs simultaneously, which asymmetrically varies the design. The design variables are initially ${d_1}={d_2}=5{m}$. Fig. \ref{ig_vib_cant_rib_mode_shapes} shows the eigenmode shapes for the lowest three eigenfrequencies. Hereafter, we denote the $i$-th eigenfrequency by ${f_i}$ which is related with the angular frequency by ${f_i} = {\omega _i}/2\pi $. The first eigenfrequency ${f_1} = 16.68Hz$ is simple, but the second and third eigenfrequencies are identical due to the symmetry in the structure and ${f_2} = {f_3} = 30.97Hz$.
Results from three different DSA methods are compared; first, we calculate the directional derivatives of eigenvalues (${\mu ^i}$) which are eigenvalues of the matrix of Eq. (\ref{ig_vib_dsa_mult_dir_deriv}). Second, we calculate the sensitivities (${\dot \zeta ^i}$) of eigenvalues regarding them as simple using Eq. (\ref{ig_vib_dsa_simple_eval_mat_deriv}) in order to see the results of this errorneous assumption. Third, in order to check the accuracy of the aforementioned two analytical DSA methods, we calculate finite difference sensitivities using a forward difference method with a perturbation amount ${10^{ - 3}}$. Table \ref{ig_vib_cant_rib_dsa_verif} shows the DSA results of the first three eigenvalues ${\zeta ^1}$, ${\zeta ^2}$, and ${\zeta ^3}$ with respect to two design variables ${d_1}$ and ${d_2}$. As the first eigenvalue is simple, both of the analytical DSA methods give the same results which agree very well with the finite difference one. Even though the second and third eigenvalues are repeated, the sensitivities ${\dot \zeta ^2}$ and ${\dot \zeta ^3}$ with respect to the design variable ${d_1}$, under the erroneous assumption of simple eigenvalues, give nearly the same values respectively with the directional derivatives ${\mu ^2}$ and ${\mu ^3}$. This is due to the symmetric design variation by changing ${d_1}$, which results in very weak off-diagonal terms in the matrix of Eq. (\ref{ig_vib_dsa_mult_dir_deriv}) \cite{seyranian1994multiple}. However, for the asymmetric design variation due to the change of ${d_2}$, the assumption of simple eigenvalue gives erroneous results while the directional derivatives agree very well with the finite difference ones. 
\begin{table*}[!htbp]
\begin{center}
\caption{DSA results of the lowest three eigenvalues (${\zeta ^i},\,\,i = 1,2,3$)}
\label{ig_vib_cant_rib_dsa_verif}
\begin{tabular}{cccccccc}
\Xhline{3\arrayrulewidth}
\multicolumn{2}{c}{\multirow{2}{*}{Mode}} & {Design} & \multirow{2}{*}{${\mu ^i}$ (a)} & \multirow{2}{*}{${\dot \zeta ^i}$ (b)} & \multirow{2}{*}{Finite difference (c)} & \multicolumn{2}{l}{Agreement (\%)} \\
\multicolumn{2}{c}{}                      & {variable \#}        &                        &                       &                                    & (a)/(c)          & (b)/(c)         \\ \Xhline{3\arrayrulewidth}
\multicolumn{2}{c}{\multirow{2}{*}{1}}    & 1                                  &9.1370E+02              &9.1370E+02             &9.1421E+02                          & 99.94            & 99.94           \\
\multicolumn{2}{c}{}                      & 2                                  &2.2842E+02              &2.2842E+02             &2.2853E+02                          & 99.95            & 99.95           \\ \hline
\multicolumn{2}{c}{\multirow{2}{*}{2}}    & 1                                  &-1.0250E+04             &-1.0249E+04            &-1.0264E+04                         & 99.86            & 99.85           \\
\multicolumn{2}{c}{}                      & 2                                  &-2.6864E+03             &-2.5496E+03            &-2.6893E+03                         & 99.89            & 94.81           \\ \hline
\multicolumn{2}{c}{\multirow{2}{*}{3}}    & 1                                  &-1.0249E+04             &-1.0250E+04            &-1.0264E+04                         & 99.85            & 99.86           \\
\multicolumn{2}{c}{}                      & 2                                  &-2.4384E+03             &-2.5752E+03            &-2.4433E+03                         & 99.80            & 105.40          \\ \Xhline{3\arrayrulewidth}
\end{tabular}
\end{center}
\end{table*}

\subsection{Hexagonal lattice on torus-shaped domain}
\begin{figure*}[htp]
\centering
\begin{minipage}{0.55\textwidth}
   \subfloat[]{\label{ig_vib_num-ex_hexagon_des_par}
      \includegraphics[width=\textwidth]{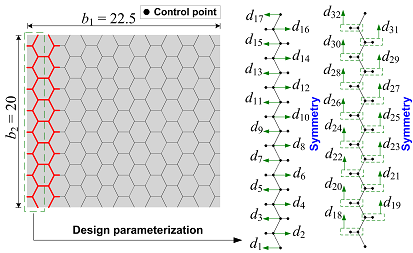}}
\end{minipage}%
\quad\quad
\begin{minipage}{0.675\textwidth}
   \subfloat[]{\label{ig_vib_num-ex_hexagon_tar_par}
      \includegraphics[width=\textwidth]{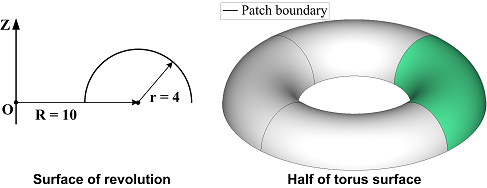}}
\end{minipage}%
   \caption{Geometric modeling of hexagonal honeycomb lattice structure on torus-shaped domain.}\label{ig_vib_hexagon_par_domain_des_par}
\end{figure*}

\begin{figure*}
\centering
  \includegraphics[width=0.9\textwidth]{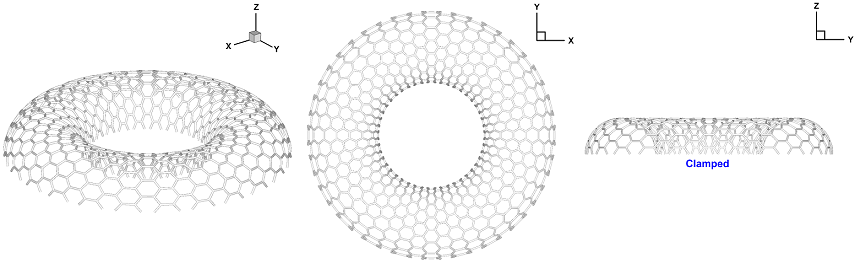}
\caption{Three different views of the original design of lattice structure.}
\label{ig_vib_num-ex_hexagon_org_des}       
\end{figure*}

We consider an optimization problem for maximizing the fundamental frequency under a volume constraint. The optimization problem can be stated as
\begin{align} 
&{\rm{Maximize}}\,\,\,\,{\psi _1} \equiv {\omega _j}^2,\label{ig_vib_num-ex_hexagon_opt_prob_obj} \\
&{\rm{subject}}\,\,{\rm{to}}\,\,\,\,{g_1} \equiv V/{V_0} - 1 \le 0, \label{ig_vib_num-ex_hexagon_opt_prob_cnst}
\end{align}
where ${\omega _j}$ denotes the angular frequency of $j$-th eigenmode, and the structural volume is calculated by 
\begin{equation} 
V \equiv \int_\Omega  {Ads}  = \sum\limits_{i = 1}^{ne} {\int_{{\Xi _i}} {A{J_c}d\xi } }.
\end{equation}
$A$ denotes the cross-sectional area. Let ${V_0}$ denote the volume of the original design, then an equality constraint $V = {V_0}$ can be implemented by employing an additional inequality constraint
\begin{equation} 
\label{ig_vib_num-ex_hexagon_vol_cnst_eq}
{g_2} \equiv  - V/{V_0} + 1 \le 0.
\end{equation}
We consider a hexagonal lattice structure shown in Fig. \ref{ig_vib_num-ex_hexagon_des_par}, where the grey region represents an initial parent domain, and the red-colored cell parameterized by linear B-splines is repeated by a translation. 32 design variables are selected as position changes of control points in the planar domain. A toroidal surface of small radius $r = 4m$ and large radius $R = 10m$ is constructed by the revolution of a half circle with a radius $r$ around the $Z$-axis, as illustrated in Fig. \ref{ig_vib_num-ex_hexagon_tar_par}, which is selected as a target parent domain, and consists of four quadratic NURBS patches. In this example, we assume that the target parent domain does not have a design dependence. The initial lattice embedded in the initial parent domain (grey-region) of Fig. \ref{ig_vib_num-ex_hexagon_des_par} is mapped into each of the four NURBS patches in the target parent domain, for example, a green-colored region of Fig. \ref{ig_vib_num-ex_hexagon_tar_par}. Fig. \ref{ig_vib_num-ex_hexagon_org_des} shows the constructed hexagonal lattice structure in three different views. We impose clamped boundary conditions at the whole bottom material points. A circular cross-section with radius $10cm$ is chosen. The material properties are selected as Young's modulus $E = 210\rm{GPa}$, Poisson's ratio $\nu  = 0.29$, mass density $\rho  = 7,850kg/{m^3}$.

\begin{figure*}[htp]
\centering
\begin{minipage}{0.27\textwidth}
   \subfloat[Initial parent domain]{\label{ig_vib_num-ex_hexagon_opt_des_init_par}
      \includegraphics[width=\textwidth]{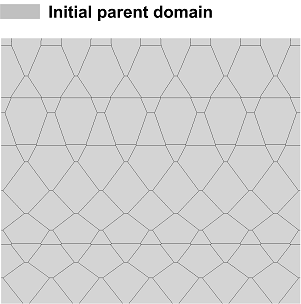}}
\end{minipage}%
\quad\quad
\begin{minipage}{0.6\textwidth}
   \subfloat[Lattice structure]{\label{ig_vib_num-ex_hexagon_opt_des_tar_par}
      \includegraphics[width=\textwidth]{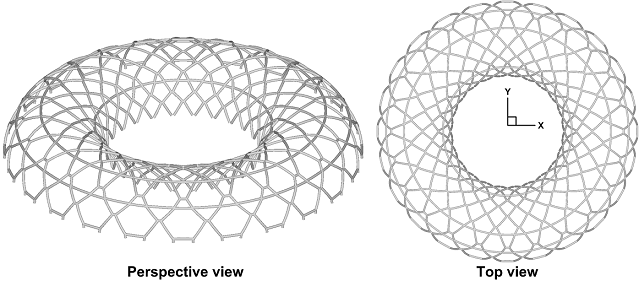}}
\end{minipage}%
   \caption{Optimal design of lattice structure}\label{ig_vib_hexagon_opt_design}
\end{figure*}

\begin{figure*}[htp]
\centering
\begin{minipage}{0.45\textwidth}
   \subfloat[Fundamental frequency]{\label{ig_vib_num-ex_hexagon_history_freq}
      \includegraphics[width=\textwidth]{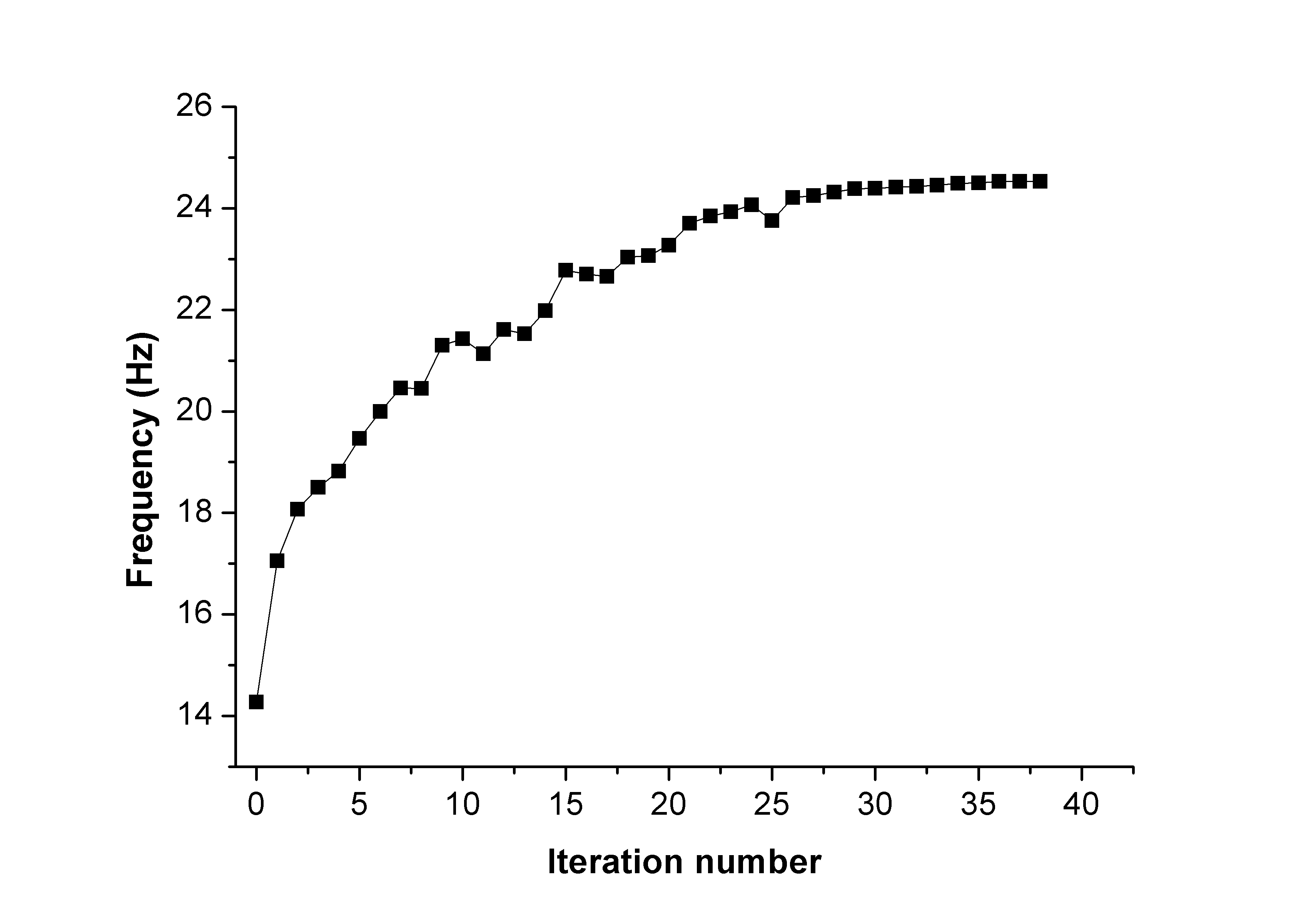}}
\end{minipage}%
\quad
\begin{minipage}{0.45\textwidth}
   \subfloat[Volume]{\label{ig_vib_num-ex_hexagon_history_vol}
      \includegraphics[width=\textwidth]{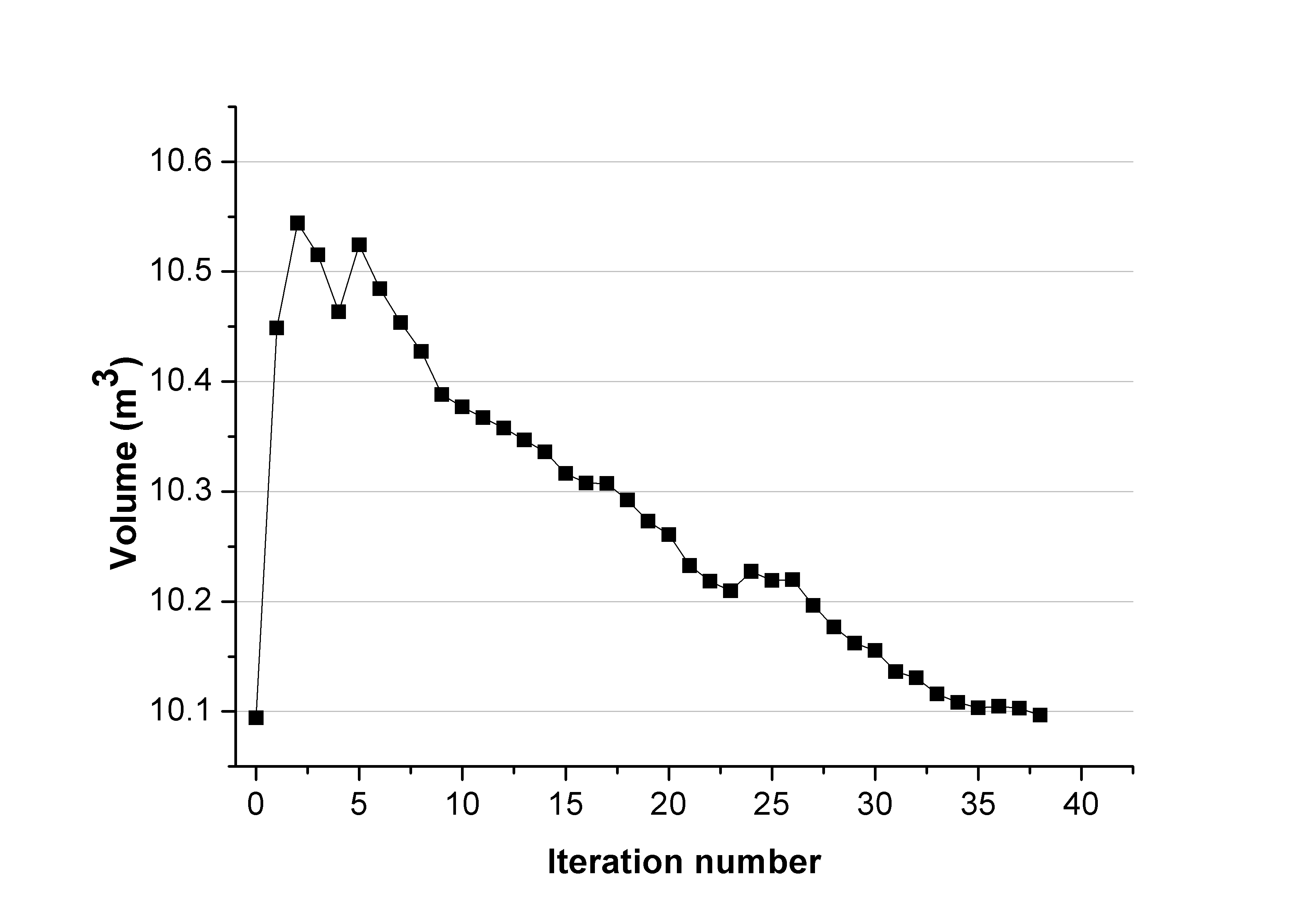}}
\end{minipage}%
   \caption{Optimization history}\label{ig_vib_hexagon_opt_history}
\end{figure*}

\begin{figure*}[htp]
\centering
\begin{minipage}{0.6\textwidth}
   \subfloat[Original design (${f_1} = 14.27{\rm{Hz}}$)]{\label{ig_vib_num-ex_hexagon_deform_org_des}
      \includegraphics[width=\textwidth]{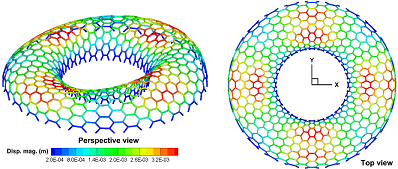}}
\end{minipage}%
\\
\begin{minipage}{0.6\textwidth}
   \subfloat[Optimal design (${f_1} = 24.53{\rm{Hz}}$)]{\label{ig_vib_num-ex_hexagon_deform_opt_des}
      \includegraphics[width=\textwidth]{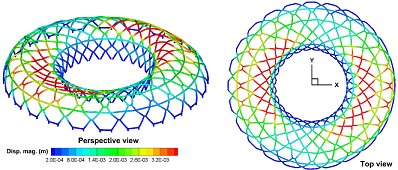}}
\end{minipage}%
   \caption{Comparison of the first eigenmode shape (scaling factor 200 used)}\label{ig_vib_hexagon_deform_opt_design}
\end{figure*}

We find an optimal design for maximizing the fundamental frequency while maintaining the same volume by solving the optimization problem of Eqs. (\ref{ig_vib_num-ex_hexagon_opt_prob_obj}), (\ref{ig_vib_num-ex_hexagon_opt_prob_cnst}), and (\ref{ig_vib_num-ex_hexagon_vol_cnst_eq}), where an SQP (Sequential Quadratic Programming) algorithm is utilized. The optimal design of the lattice structure is shown in Fig. \ref{ig_vib_hexagon_opt_design}. Fig. \ref{ig_vib_hexagon_opt_history} shows the optimization history where the fundamental frequency increases from 14.27Hz to 24.53Hz, while the volume remains nearly the same within a specified tolerance of constraint violation. Fig. \ref{ig_vib_hexagon_deform_opt_design} compares the deformed configurations of fundamental eigenmodes in the original and optimal designs.

\subsection{Jacket tower model}
\subsubsection{Modeling of constrained straight beam structures}
\label{modeling_linear}
\begin{figure*}[htp]
\centering
\begin{minipage}{0.25\textwidth}
   \subfloat[Initial parent domain]{\label{ig_vib_num-ex_tower_init_par}
      \includegraphics[width=\textwidth]{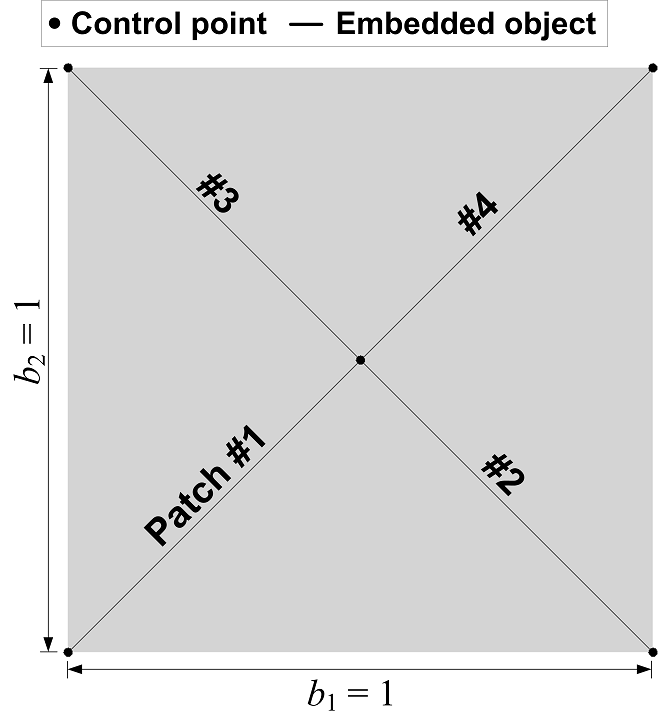}}
\end{minipage}%
\quad\quad
\begin{minipage}{0.25\textwidth}
   \subfloat[Target parent domain]{\label{ig_vib_num-ex_tower_tar_par}
      \includegraphics[width=\textwidth]{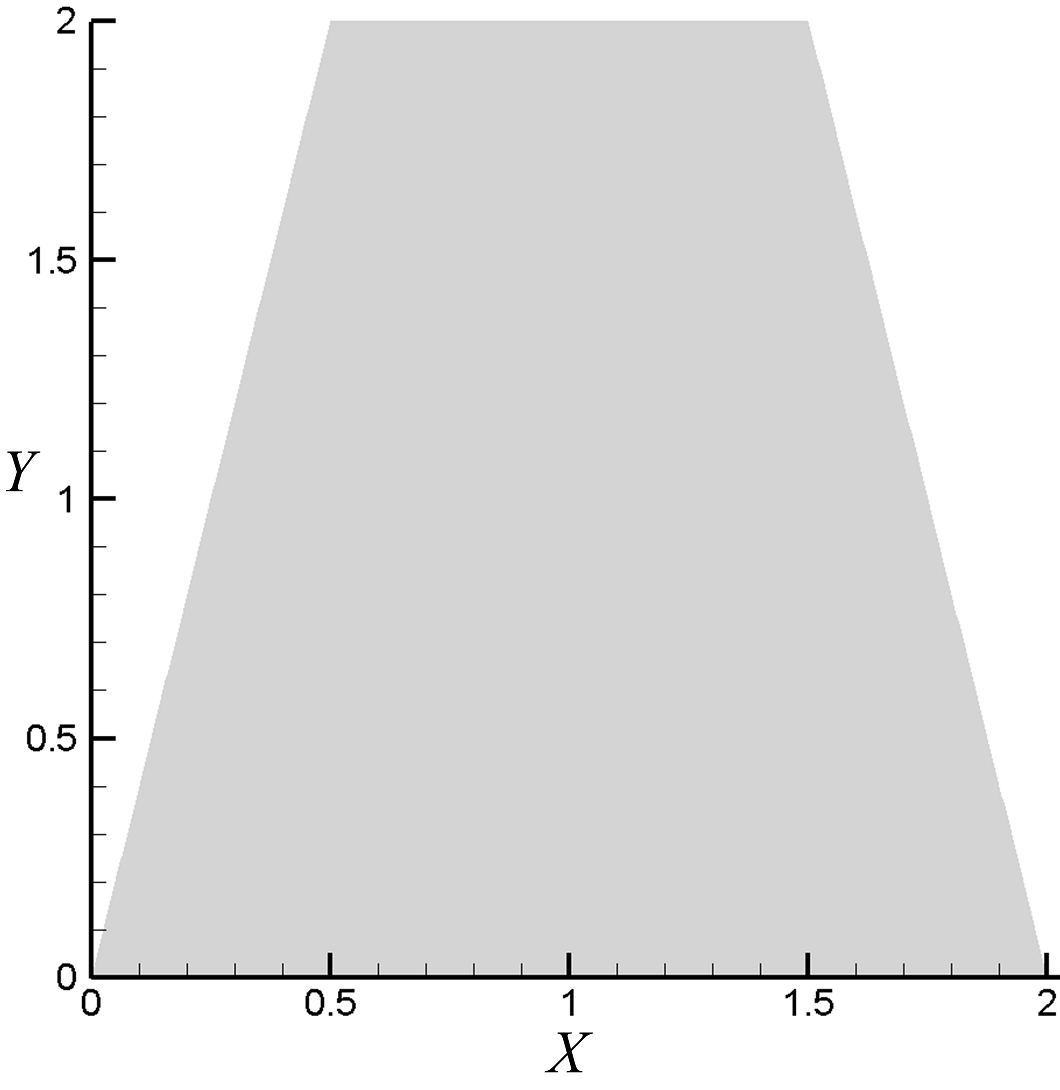}}
\end{minipage}%
   \caption{A modeling of curve geometry on a trapezoidal domain}\label{ig_vib_tower_model}
\end{figure*}

\begin{figure*}[htp]
\centering
\begin{minipage}{0.25\textwidth}
   \subfloat[Initial parent domain]{\label{ig_vib_num-ex_tower_init_par_cp}
      \includegraphics[width=\textwidth]{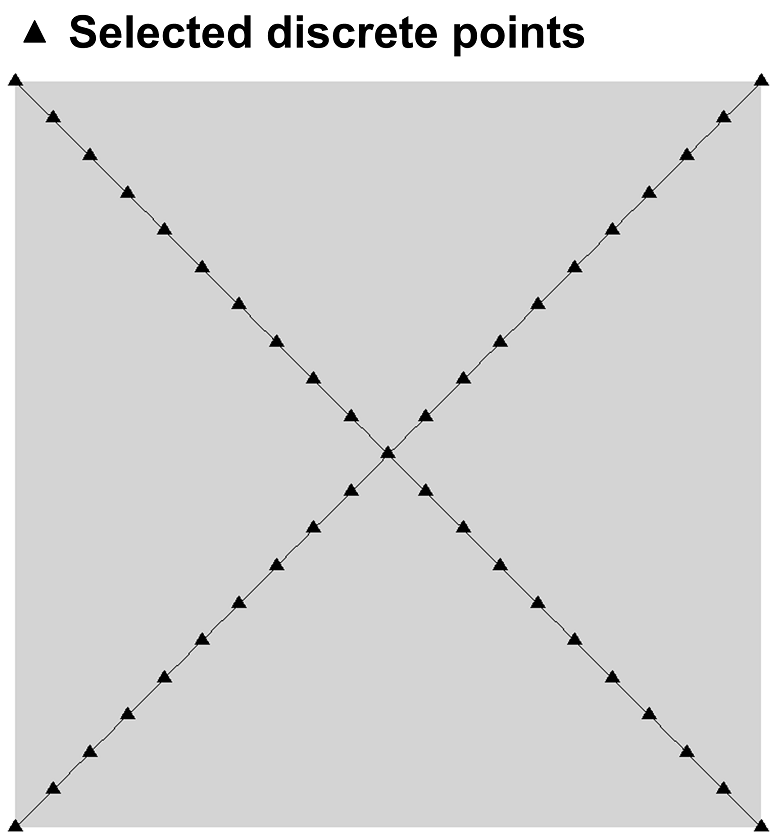}}
\end{minipage}%
\quad\quad
\begin{minipage}{0.25\textwidth}
   \subfloat[Target parent domain]{\label{ig_vib_num-ex_tower_tar_par_cp}
      \includegraphics[width=\textwidth]{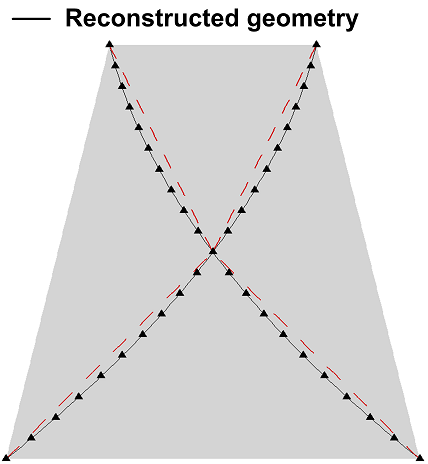}}
\end{minipage}%
   \caption{Curved geometry of embedded object due to a nonlinear mapping between parent domains}\label{ig_vib_num-ex_tower_par_cp}
\end{figure*}

\begin{figure*}[htp]
\centering
\begin{minipage}{0.25\textwidth}
   \subfloat[Initial parent domain]{\label{ig_vib_num-ex_tower_init_par_end_pt}
      \includegraphics[width=\textwidth]{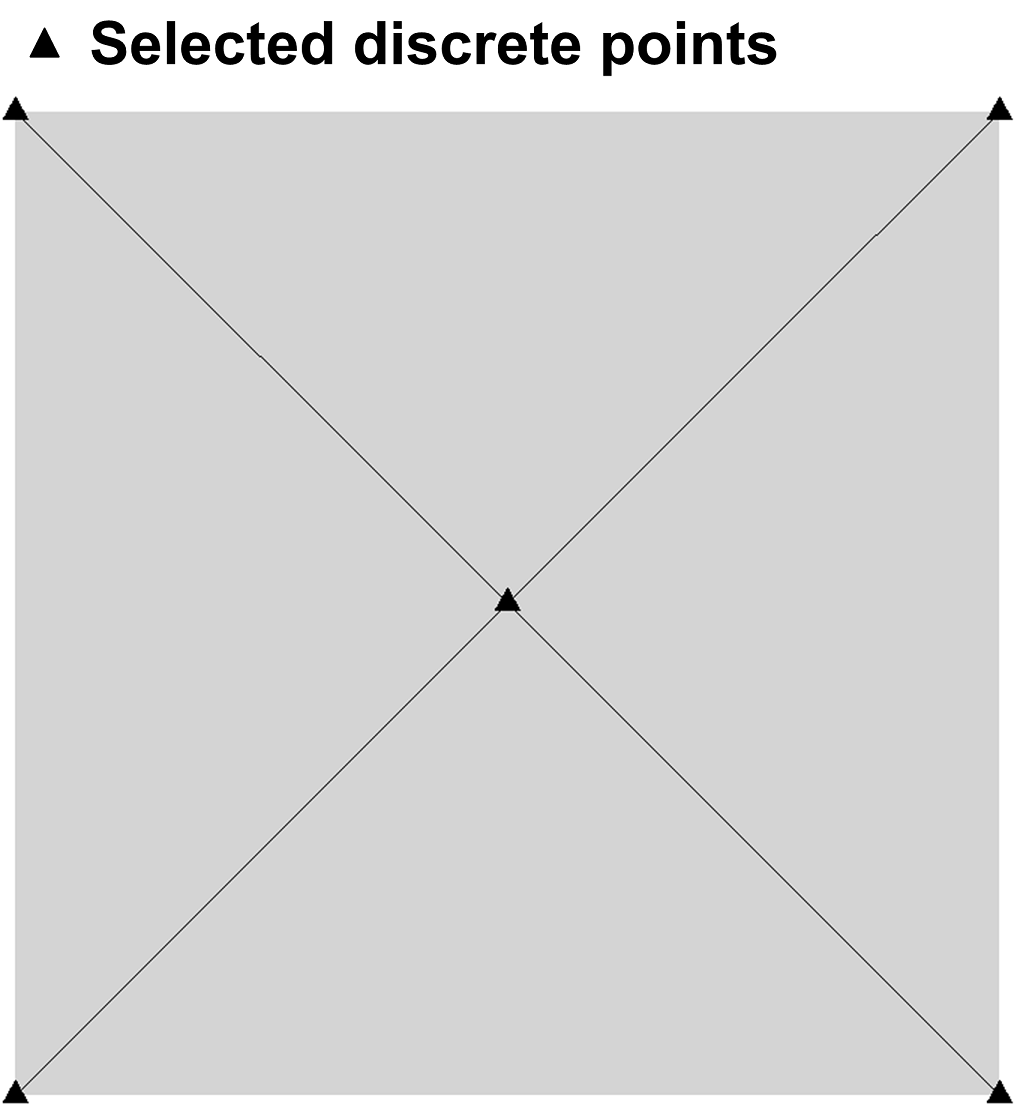}}
\end{minipage}%
\quad\quad
\begin{minipage}{0.25\textwidth}
   \subfloat[Target parent domain]{\label{ig_vib_num-ex_tower_tar_par_end_pt}
      \includegraphics[width=\textwidth]{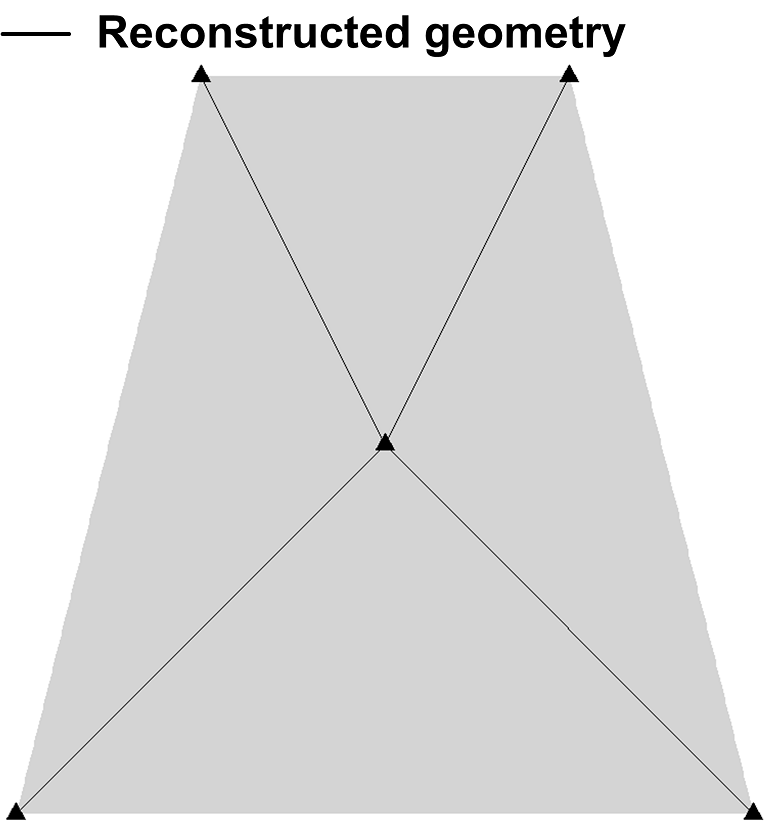}}
\end{minipage}%
   \caption{A modeling of straight curve geometry on a trapezoidal domain}\label{ig_vib_tower_model_end_point}
\end{figure*}

For a constrained structure on a planar surface, straight geometries of structural members are often preferred for their manufacturability. However, even though a target parent domain is a planar surface, the FFD process sometimes does not provide a straight curve geometry. To illustrate this issue and a solution, we consider the geometric modeling of the curves embedded in a trapezoidal area. An initial parent domain has a square shape with a side length ${b_1} = {b_2} = 1$, and an embedded structure is modeled by four single element linear B-spline patches, as shown in Fig. \ref{ig_vib_num-ex_tower_init_par}. A target parent domain is chosen as a trapezoidal domain shown in Fig. \ref{ig_vib_num-ex_tower_tar_par}.
For each of the four linear B-spline curves in the initial parent domain, 11 equidistant points are selected, as indicated by triangular dots in Fig. \ref{ig_vib_num-ex_tower_init_par_cp}. Those points are mapped onto the target parent domain, and they are not linearly aligned; thus, the reconstruction of each patch by a global curve interpolation of the 11 discrete points does not provide a straight geometry, that is, the reconstructed curves deviate from the desired ones indicated by dotted lines, as illustrated in Fig. \ref{ig_vib_num-ex_tower_tar_par_cp}. For more detailed discussion of a nonlinear mapping between the initial and target parent domains, see Appendix A. In order to have a straight curve, it is required to use a linear interpolation within each patch. For each of the four curves in the initial parent domain, we select two end points indicated by triangular dots in Fig. \ref{ig_vib_num-ex_tower_init_par_end_pt}. Those points are mapped onto the target parent domain, and are linearly interpolated so that a straight reconstructed geometry is obtained, as shown in Fig. \ref{ig_vib_num-ex_tower_tar_par_end_pt}. These straight B-spline curves can be more refined by a degree elevation or a knot insertion process for more accurate response analysis and DSA, and it should be noted that those refinement processes do not alter the curves either geometrically or parametrically.
\subsubsection{Design optimization}
\begin{figure*}[htp]
\centering
\begin{minipage}{0.25\textwidth}
   \subfloat[]{\label{ig_vib_num-ex_tower_dv}
      \includegraphics[width=0.8\textwidth]{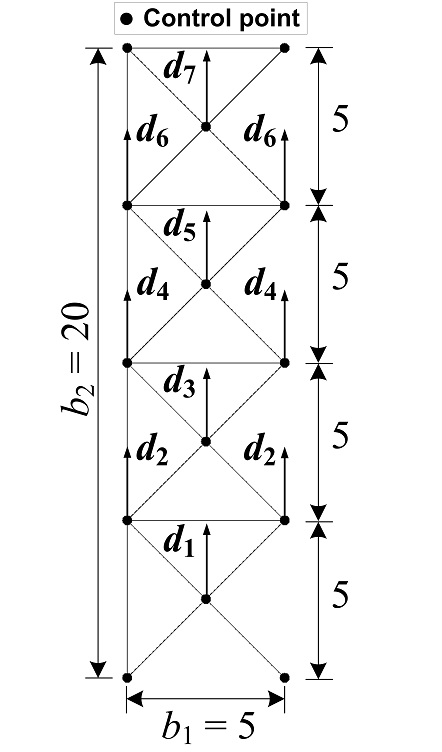}}
\end{minipage}%
\quad\quad
\begin{minipage}{0.3325\textwidth}
   \subfloat[]{\label{ig_vib_num-ex_tower_tar_par_dv}
      \includegraphics[width=0.8\textwidth]{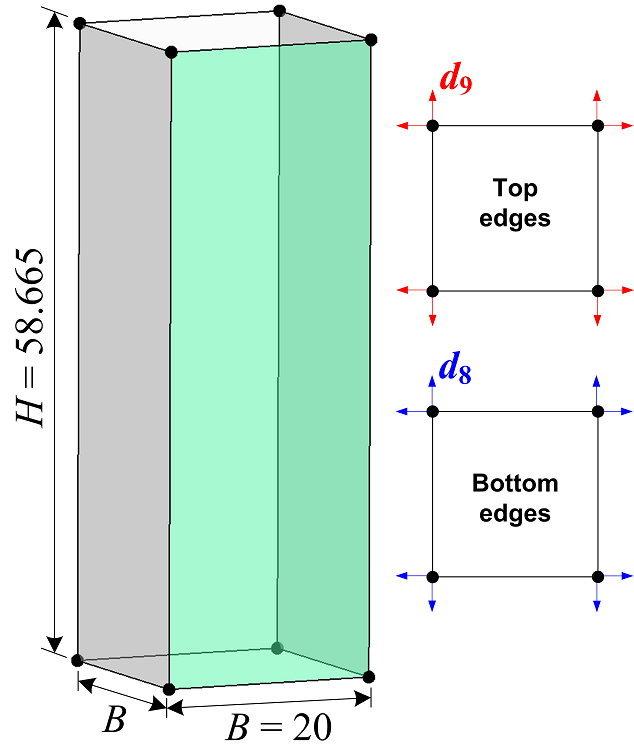}}
\end{minipage}%
\quad\quad
\begin{minipage}{0.2\textwidth}
   \subfloat[]{\label{ig_vib_num-ex_tower_org_des}
      \includegraphics[width=0.8\textwidth]{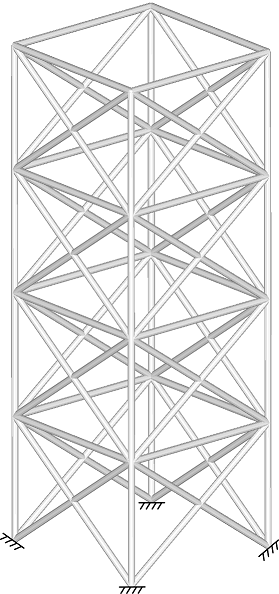}}
\end{minipage}%
   \caption{Geometric modeling and design variables of jacket tower structure. (a) Initial parent domain, (b) Target parent domain, (c) Jacket tower}\label{ig_vib_tower_model_dv_org_des}
\end{figure*}

\begin{figure*}[htp]
\centering
\begin{minipage}{0.22\textwidth}
   \subfloat[Initial parent domain]{\label{ig_vib_num-ex_tower_opt_des_init_par}
      \includegraphics[width=\textwidth]{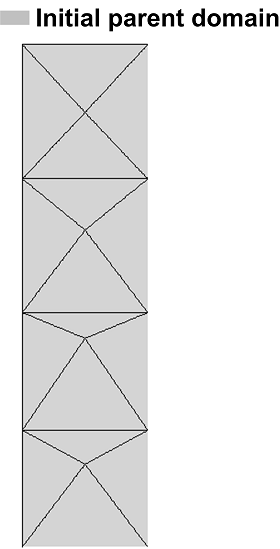}}
\end{minipage}%
\quad\quad
\begin{minipage}{0.45\textwidth}
   \subfloat[Jacket tower structure]{\label{ig_vib_num-ex_tower_opt_des_tar_par}
      \includegraphics[width=\textwidth]{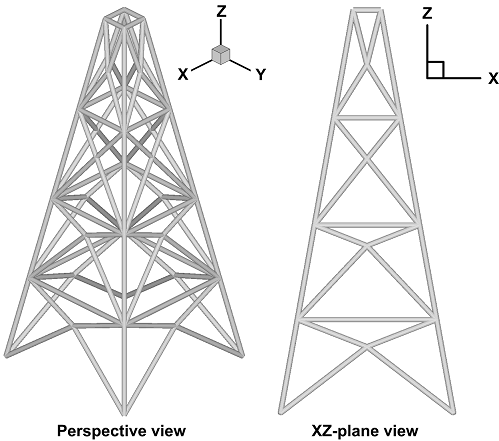}}
\end{minipage}%
   \caption{Optimal design of jacket tower structure}\label{ig_vib_tower_model_opt_des}
\end{figure*}

\begin{figure*}[htp]
\centering
\begin{minipage}{0.475\textwidth}
   \subfloat[Fundamental frequency]{\label{ig_vib_num-ex_tower_opt_history_freq}
      \includegraphics[width=\textwidth]{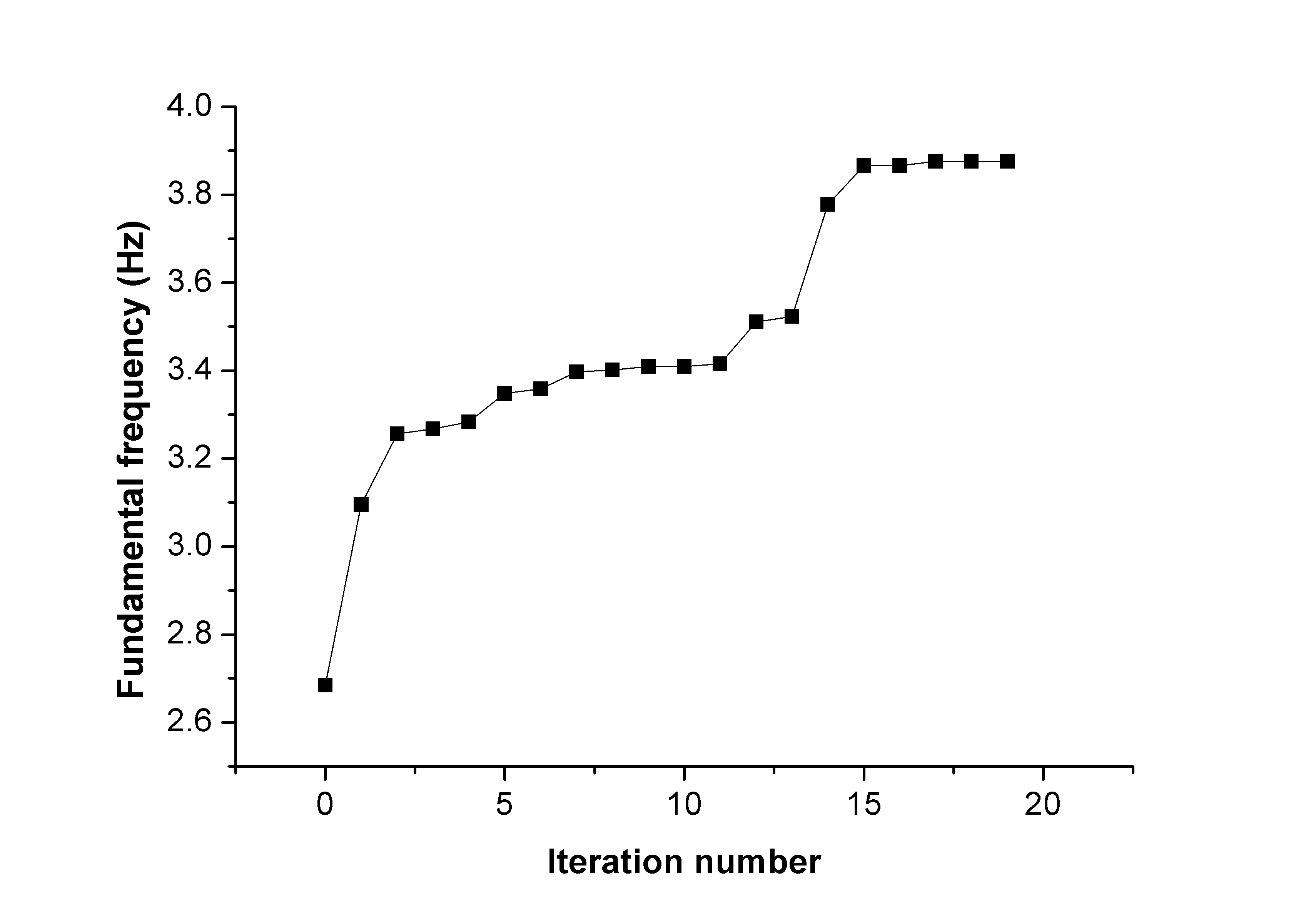}}
\end{minipage}%
\quad
\begin{minipage}{0.475\textwidth}
   \subfloat[Volume]{\label{ig_vib_num-ex_tower_opt_history_vol}
      \includegraphics[width=\textwidth]{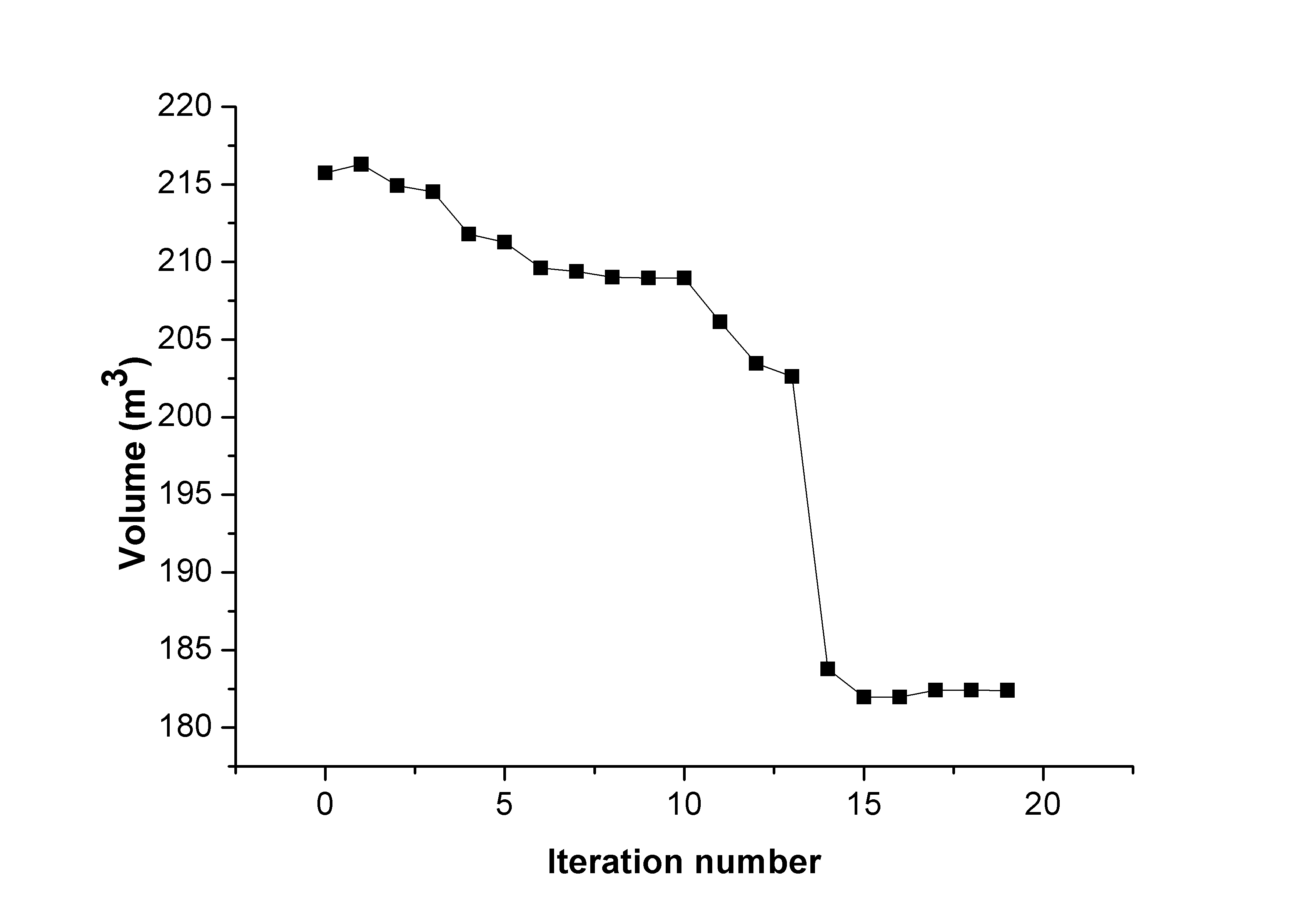}}
\end{minipage}%
   \caption{Optimization history}\label{ig_vib_tower_opt_history}
\end{figure*}

\begin{figure*}[htp]
\centering
\begin{minipage}{0.3175\textwidth}
   \subfloat[Original design (${f_1} = 2.68{\rm{Hz}}$)]{\label{ig_vib_num-ex_tower_org_des_deform}
      \includegraphics[width=\textwidth]{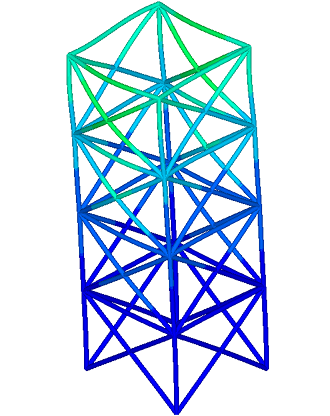}}
\end{minipage}%
\quad\quad
\begin{minipage}{0.35\textwidth}
   \subfloat[Optimal design (${f_1} = 3.88{\rm{Hz}}$)]{\label{ig_vib_num-ex_tower_opt_des_deform}
      \includegraphics[width=\textwidth]{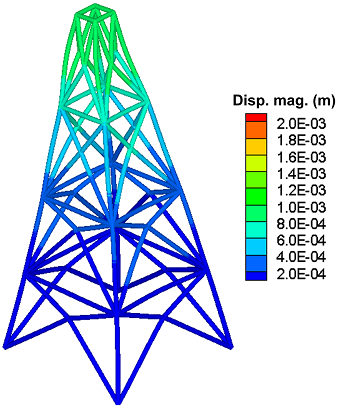}}
\end{minipage}%
   \caption{Comparison of the first mode shapes of the original and optimal designs (scaling factor 5,000 used)}\label{ig_vib_tower_first_mode_shape_deform}
\end{figure*}

In this example, we deal with a jacket tower structure, which is modeled in a way that a lattice structure shown in Fig. \ref{ig_vib_num-ex_tower_dv} embedded in a rectangular domain having width ${b_1} = 5$ and height ${b_2} = 10$ is mapped into each of the four surfaces comprising the prismatic target parent domain shown in Fig. \ref{ig_vib_num-ex_tower_tar_par_dv}, for example, the green-colored one. The target parent domain has a prismatic shape with height $H=58.665m$ and width $B=20m$. As illustrated in Fig. \ref{ig_vib_num-ex_tower_dv}, 7 configuration design variables are selected as control point positions of curves embedded in the initial parent domain. The other two design variables, indicated in Fig. \ref{ig_vib_num-ex_tower_tar_par_dv}, change the positions of top and bottom vertices in the prismatic domain while maintaining the height and the flat geometry.
We seek an optimal design having straight members; thus, as investigated in section \ref{modeling_linear}, a global curve interpolation process is conducted using a single linear B-spline element for each patch of the embedded curves. Fig. \ref{ig_vib_num-ex_tower_org_des} shows the obtained jacket tower structure, and we consider clamped boundary conditions at the bottom ends. The material properties are selected as Young's modulus $E = 210\rm{GPa}$, Poisson's ratio $\nu  = 0.29$, and mass density $\rho  = 7850kg/{m^3}$. The cross-section has a circular shape with radius $40cm$. Using the $k$-refinement strategy in IGA the simulation model is refined into 20 quartic B-spline elements in each patch. MMFD (Modified Method of Feasible Direction) optimization algorithm is used. In the optimal design shown in Fig. \ref{ig_vib_num-ex_tower_opt_des_tar_par}, it is noticeable that, the bottom dimension increases while the upper dimension decreases, which results in prismatic designs typically found in many jacket tower structures. Fig. \ref{ig_vib_num-ex_tower_opt_history_freq} shows that fundamental frequency increases from 2.68Hz in the original design to 3.88Hz in the optimal design. Meanwhile, the volume decreases, as shown in Fig. \ref{ig_vib_num-ex_tower_opt_history_vol}. Fig. \ref{ig_vib_tower_first_mode_shape_deform} compares mode shapes for the original and optimal designs.
\subsection{Twisted tower model}
\begin{figure*}[htp]
\centering
\begin{minipage}{0.325\textwidth}
   \subfloat[Initial parent domain]{\label{ig_vib_cyl_tower_init_par}
      \includegraphics[width=\textwidth]{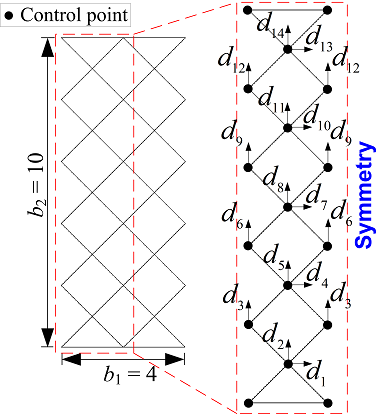}}
\end{minipage}%
\quad\quad
\begin{minipage}{0.35\textwidth}
   \subfloat[Target parent domain]{\label{ig_vib_cyl_tower_tar_par}
      \includegraphics[width=\textwidth]{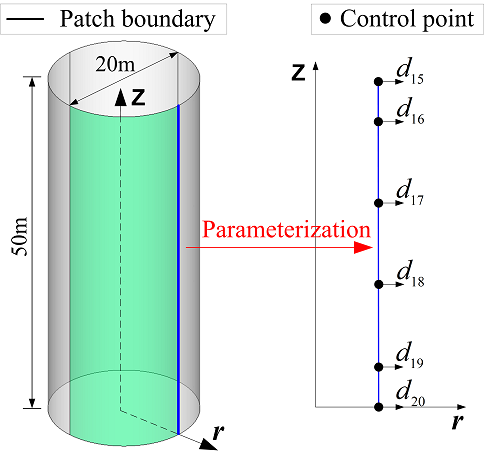}}
\end{minipage}%
   \caption{Geometric modeling and design variables of cylindrical tower structure}\label{ig_vib_cyl_tower_model_dv}
\end{figure*}

\begin{figure*}[htp]
\centering
\begin{minipage}{0.15\textwidth}
   \subfloat[Case {\#}1]{\label{ig_vib_cyl_tower_one_clamp}
      \includegraphics[width=\textwidth]{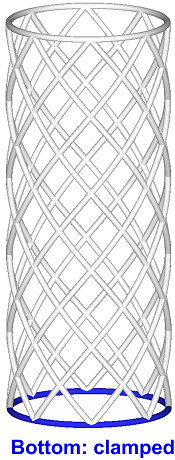}}
\end{minipage}%
\quad\quad
\begin{minipage}{0.15\textwidth}
   \subfloat[Case {\#}2]{\label{ig_vib_cyl_tower_two_clamp}
      \includegraphics[width=\textwidth]{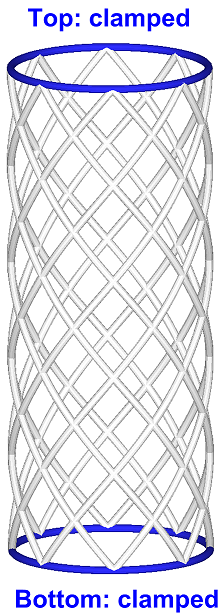}}
\end{minipage}%
   \caption{Lattice structure with two cases of boundary conditions}\label{ig_vib_cyl_tower_two_bdc}
\end{figure*}

\begin{figure*}[htp]
\centering
\begin{minipage}{0.25\textwidth}
   \subfloat[Initial parent domain]{\label{ig_vib_cyl_tower_one_clamp}
      \includegraphics[width=\textwidth]{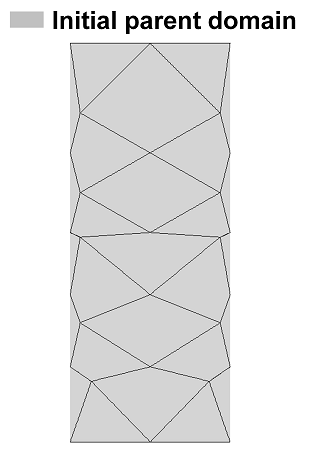}}
\end{minipage}%
\quad\quad
\begin{minipage}{0.45\textwidth}
   \subfloat[Twisted tower]{\label{ig_vib_cyl_tower_two_clamp}
      \includegraphics[width=\textwidth]{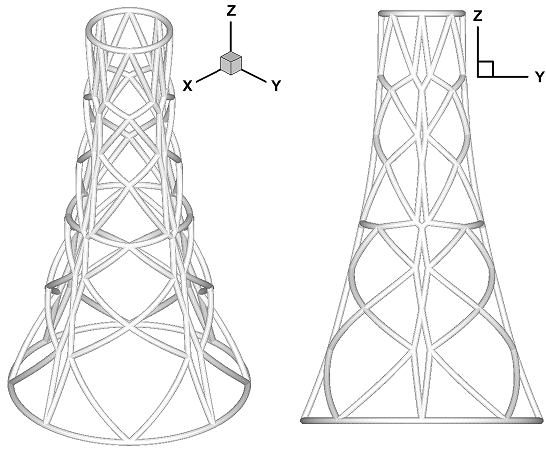}}
\end{minipage}%
   \caption{Optimal design for the boundary condition case {\#}1 (bottom clamped)}\label{ig_vib_cyl_tower_opt_des}
\end{figure*}

\begin{figure*}[htp]
\centering
\begin{minipage}{0.285\textwidth}
   \subfloat[Initial parent domain]{\label{ig_vib_cyl_tower_one_clamp}
      \includegraphics[width=\textwidth]{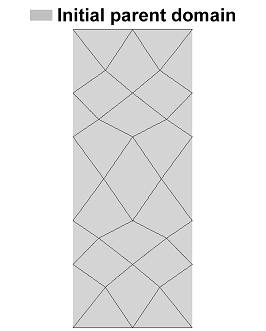}}
\end{minipage}%
\quad\quad
\begin{minipage}{0.45\textwidth}
   \subfloat[Twisted tower]{\label{ig_vib_cyl_tower_two_clamp}
      \includegraphics[width=\textwidth]{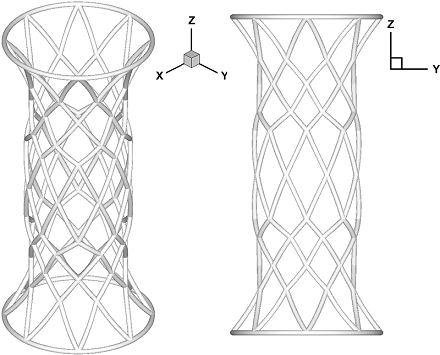}}
\end{minipage}%
   \caption{Optimal design for the boundary condition case {\#}2 (top and bottom clamped)}\label{ig_vib_cyl_tower_opt2_des}
\end{figure*}

\begin{figure*}[htp]
\centering
\begin{minipage}{0.485\textwidth}
   \subfloat[Fundamental frequency]{\label{ig_vib_cyl_tower_opt_history_freq}
      \includegraphics[width=\textwidth]{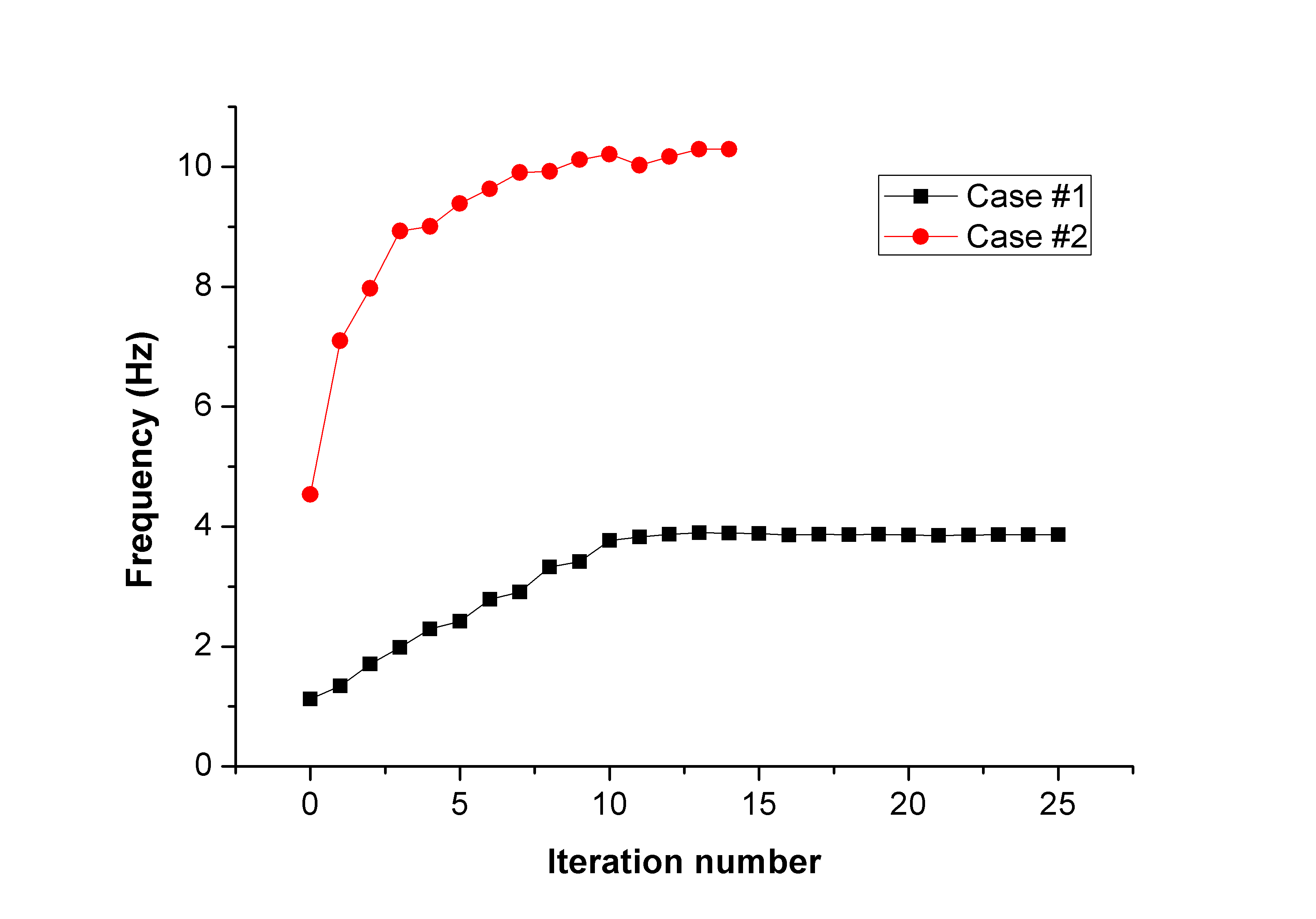}}
\end{minipage}%
\quad
\begin{minipage}{0.485\textwidth}
   \subfloat[Volume]{\label{ig_vib_cyl_tower_opt_history_vol}
      \includegraphics[width=\textwidth]{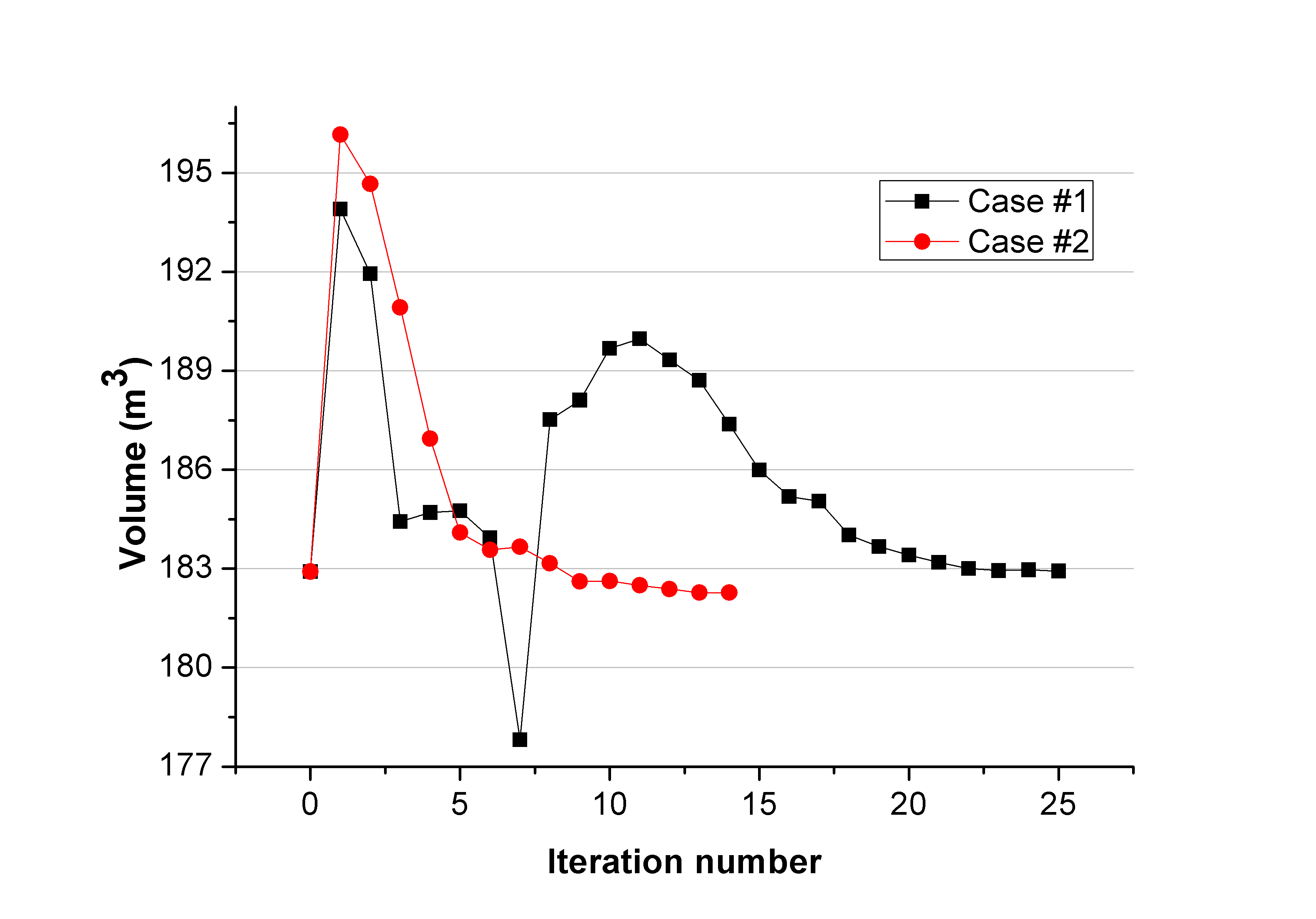}}
\end{minipage}%
   \caption{Optimization history}\label{ig_vib_cyl_tower_opt12_history}
\end{figure*}

\begin{figure*}[htp]
\centering
\begin{minipage}{0.275\textwidth}
   \subfloat[Original design (${f_1} = 1.13Hz$)]{\label{ig_vib_cyl_tower_deform_one_fix_org_des}
      \includegraphics[width=\textwidth]{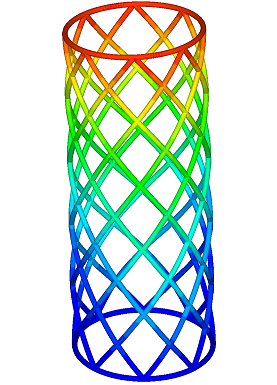}}
\end{minipage}%
\quad\quad
\begin{minipage}{0.3\textwidth}
   \subfloat[Optimal design (${f_1} = 3.86Hz$)]{\label{ig_vib_cyl_tower_deform_one_fix_opt_des}
      \includegraphics[width=\textwidth]{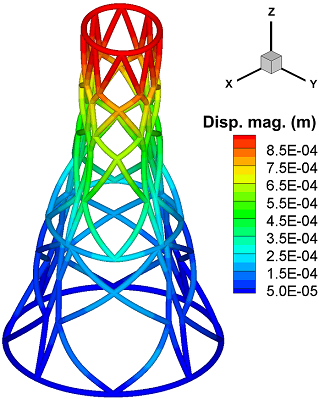}}
\end{minipage}%
   \caption{Comparison of the first eigenmode shapes for case {\#}1 (scaling factor 1,000 used)}\label{ig_vib_cyl_tower_deform_one_fix}
\end{figure*}

\begin{figure*}[htp]
\centering
\begin{minipage}{0.325\textwidth}
   \subfloat[Original design (${f_1} = 4.53Hz$)]{\label{ig_vib_cyl_tower_deform_two_fix_org_des}
      \includegraphics[width=\textwidth]{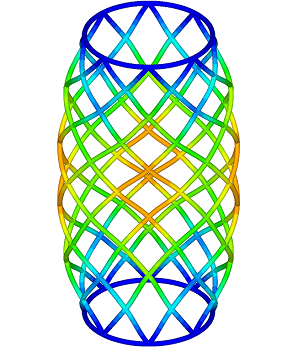}}
\end{minipage}%
\quad\quad
\begin{minipage}{0.275\textwidth}
   \subfloat[Optimal design (${f_1} = 10.30Hz$)]{\label{ig_vib_cyl_tower_deform_two_fix_opt_des}
      \includegraphics[width=\textwidth]{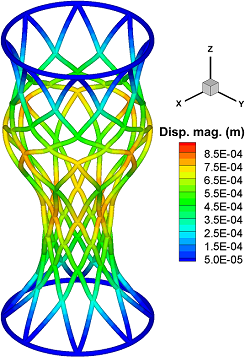}}
\end{minipage}%
   \caption{Comparison of the first eigenmode shapes for case {\#}2 (scaling factor 1,000 used)}\label{ig_vib_cyl_tower_deform_two_fix}
\end{figure*}

A rectangular domain (grey region) having width ${b_1}=4$ and height ${b_2}=10$ is considered as an initial parent domain, where a rhombic lattice structure shown in Fig. \ref{ig_vib_cyl_tower_init_par} is modeled by linear B-spline curves, and it has 14 design variables ${d_1}\sim{d_{14}}$ as changes of control point positions. As illustrated in Fig. \ref{ig_vib_cyl_tower_tar_par}, a target parent domain is a cylindrical surface which consists of four NURBS patches, and is constructed by a surface of revolution of a straight line parameterized by 6 control points with quadratic B-spline basis functions. The design of target parent domain is parameterized by 6 design variables ${d_{15}}\sim{d_{20}}$; thus, total 20 configuration design variables are employed in this example. We map the rhombic lattice to each of the four NURBS patches of the cylindrical surface, for example, the green-colored one. Then, we finally have the lattice structure shown in Fig. \ref{ig_vib_cyl_tower_two_bdc}, where two kinds of boundary conditions are considered; first, bottom members are fixed (case {\#}1), and second, the bottom and top members are fixed (case {\#}2). The material properties are selected as Young's modulus $E = 210\rm{GPa}$, Poisson's ratio $\nu  = 0.29$, mass density $\rho  = 7,850kg/{m^3}$. The cross-section has a circular shape with radius 40cm.
We seek a structural configuration having maximal fundamental frequency under a volume constraint, that is, the optimization problem of Eqs. (\ref{ig_vib_num-ex_hexagon_opt_prob_obj}) and (\ref{ig_vib_num-ex_hexagon_opt_prob_cnst}) is solved, where a SQP algorithm is used. Figs. \ref{ig_vib_cyl_tower_opt_des} and \ref{ig_vib_cyl_tower_opt2_des} illustrate the optimal designs, where the changes of overall structural shapes into tapered designs are noticeable. Fig. \ref{ig_vib_cyl_tower_opt_history_freq} shows the change of fundamental frequency during the optimization process. At the original design, the case {\#}2 shows a larger frequency than that of the case {\#}1 since it has larger stiffness due to more kinematic constraints. The optimal designs with an increased radius of the cylindrical domain in clamped regions have larger stiffness while the structural volume is nearly maintained after the optimizations in both cases (see Fig. \ref{ig_vib_cyl_tower_opt_history_vol}), so that fundamental frequencies increase significantly through the optimizations from 1.13Hz and 4.53Hz in the initial design to 3.86Hz and 10.30Hz in the optimized designs for the cases {\#}1 and {\#}2, respectively. Figs. \ref{ig_vib_cyl_tower_deform_one_fix} and \ref{ig_vib_cyl_tower_deform_two_fix} compare the mode shapes of the first eigenmode in the original and optimal designs. 
\section{Conclusions}
In this paper, we present a gradient-based optimal configuration design of three-dimensional built-up beam structures for maximizing fundamental frequencies. A shear-deformable beam model is employed for the analyses of structural vibrations within an isogeometric analysis framework using the NURBS basis functions. We particularly extend the previous work \cite{choi2018constrained} in the perspective of considering a configuration design dependence of a curved surface where beam structures are embedded. This gives more design degrees-of-freedom, and provides an effective scheme for design parameterization and design velocity field computation in three-dimensional beam structures constrained on a curved surface. We also derive an analytical configuration DSA expressions for repeated eigenvalues. The developed design sensitivity analysis and optimization method is demonstrated through various numerical examples. 

\section*{Acknowledgements}
\label{Acknowledgements}
\noindent M.-J. Choi and B. Koo were supported by the National Research Foundation of Korea (NRF) grant funded by the Korea government (Ministry of Science, ICT, and Future Planning) (No. NRF-2018R1D1A1B07050370)

\section*{Conflict of interest}
The authors declare that they have no conflict of interest.

\section*{Replication of results}
All the expressions are implemented using FORTRAN and the resulting data are processed by Tecplot. The processed data used to support the findings of this study are available from the corresponding author upon request. If the information provided in the paper is not sufficient, interested readers are welcome to contact the authors for further explanations.

\renewcommand{\theequation}{A.\arabic{equation}}
\setcounter{equation}{0}  
\section*{Appendix}
\label{vib_app_a}
In this appendix, we analytically illustrate the FFD process may yield a nonlinear mapping of embedded curves even if the target parent domain is a linear B-spline surface. We assume both of the square and trapezoidal domains respectively for initial and target parent domains, shown in Fig. \ref{ig_vib_tower_model}, are composed of a single linear B-spline surface element, and the basis functions are given by
\begin{equation} \label{ig_vib_app_basis}
\left. \begin{array}{lcl}
{{\tilde W}_1} &=& (1 - {{\tilde \xi }^1})(1 - {{\tilde \xi }^2})\\
{{\tilde W}_2} &=& {{\tilde \xi }^1}(1 - {{\tilde \xi }^2})\\
{{\tilde W}_3} &=& (1 - {{\tilde \xi }^1}){{\tilde \xi }^2}\\
{{\tilde W}_4} &=& {{\tilde \xi }^1}{{\tilde \xi }^2}
\end{array} \right\},
\end{equation}
where the corresponding control point positions are ${\bf{\tilde B}}_1^0(0,0)$, ${\bf{\tilde B}}_2^0(1,0)$, ${\bf{\tilde B}}_3^0(0,1)$, and ${\bf{\tilde B}}_4^0(1,1)$ for the initial parent domain, and ${{\bf{\tilde B}}_1}(0,0)$, ${\bf{\tilde B}}_2^0(2,0)$, ${\bf{\tilde B}}_3^0(0.5,2)$, and   ${\bf{\tilde B}}_4^0(1.5,2)$ for the target parent domain. The target parent domain is expressed by a linear combination of the basis functions of Eq. (\ref{ig_vib_app_basis}) and the control point positions ${{\bf{\tilde B}}_i}$ ($i=1,...,4$) through Eq. (\ref{ig_vib_embed_curve_tar_par}), as
\begin{equation} \label{ig_vib_app_x_tilde}
{\bf{\tilde X}}({\tilde \xi ^1},{\tilde \xi ^2}) = \left[ {\begin{array}{*{20}{c}}
{2{{\tilde \xi }^1} + 0.5{{\tilde \xi }^2} + 0.3{{\tilde \xi }^1}{{\tilde \xi }^2}}\\
{2{{\tilde \xi }^2}}
\end{array}} \right].
\end{equation}
Using Eq. (\ref{ig_vib_embed_curve_param_pos_algebra}) and ${b_1} = {b_2} = 1$, the surface parametric coordinates corresponding to the curve parametric position is obtained by
\begin{equation} \label{ig_vib_app_xi_tilde_12}
\left. \begin{array}{c}
{{\tilde \xi }^1} = X_1^0\\
{{\tilde \xi }^2} = X_2^0
\end{array} \right\}.
\end{equation}
Consequently, substituting Eq. (\ref{ig_vib_app_xi_tilde_12}) into Eq. (\ref{ig_vib_app_x_tilde}) yields
\begin{equation} \label{ig_vib_app_x_tilde_phys}
{\bf{\tilde X}}(X_1^0,X_2^0) = \left[ {\begin{array}{*{20}{c}}
{2X_1^0 + 0.5X_2^0 + 0.3X_1^0X_2^0}\\
{2X_2^0}
\end{array}} \right],
\end{equation}
which shows that the linear curve in the initial parent domain is mapped into a quadratic curve in the target parent domain.

\bibliographystyle{spmpsci}      
\bibliography{main}   

\newcommand{\noopsort}[1]{} \newcommand{\printfirst}[2]{#1}
  \newcommand{\singleletter}[1]{#1} \newcommand{\switchargs}[2]{#2#1}
\begin{thebibliography}{10}
\providecommand{\url}[1]{{#1}}
\providecommand{\urlprefix}{URL }
\expandafter\ifx\csname urlstyle\endcsname\relax
  \providecommand{\doi}[1]{DOI~\discretionary{}{}{}#1}\else
  \providecommand{\doi}{DOI~\discretionary{}{}{}\begingroup
  \urlstyle{rm}\Url}\fi

\bibitem{allaire2005level}
Allaire, G., Jouve, F.: A level-set method for vibration and multiple loads
  structural optimization.
\newblock Computer methods in applied mechanics and engineering
  \textbf{194}(30-33), 3269--3290 (2005)

\bibitem{blom2008design}
Blom, A.W., Setoodeh, S., Hol, J.M., G{\"u}rdal, Z.: Design of
  variable-stiffness conical shells for maximum fundamental eigenfrequency.
\newblock Computers \& Structures \textbf{86}(9), 870--878 (2008)

\bibitem{choi2018constrained}
Choi, M.J., Cho, S.: Constrained isogeometric design optimization of lattice
  structures on curved surfaces: computation of design velocity field.
\newblock Structural and Multidisciplinary Optimization \textbf{58}(1), 17--34
  (2018)

\bibitem{choi2019isogeometric}
Choi, M.J., Cho, S.: Isogeometric configuration design sensitivity analysis of
  geometrically exact shear-deformable beam structures.
\newblock Computer Methods in Applied Mechanics and Engineering \textbf{351},
  153--183 (2019)

\bibitem{choi2019optimal}
Choi, M.J., Oh, M.H., Koo, B., Cho, S.: Optimal design of lattice structures
  for controllable extremal band gaps.
\newblock Scientific reports \textbf{9}(1), 9976 (2019)

\bibitem{choi2016isogeometric}
Choi, M.J., Yoon, M., Cho, S.: Isogeometric configuration design sensitivity
  analysis of finite deformation curved beam structures using jaumann strain
  formulation.
\newblock Computer Methods in Applied Mechanics and Engineering \textbf{309},
  41--73 (2016)

\bibitem{du2007topological}
Du, J., Olhoff, N.: Topological design of freely vibrating continuum structures
  for maximum values of simple and multiple eigenfrequencies and frequency
  gaps.
\newblock Structural and Multidisciplinary Optimization \textbf{34}(2), 91--110
  (2007)

\bibitem{goldstein2002classical}
Goldstein, H., Poole, C., Safko, J.: Classical mechanics (2002)

\bibitem{ha2006design}
Ha, Y., Cho, S.: Design sensitivity analysis and topology optimization of
  eigenvalue problems for piezoelectric resonators.
\newblock Smart Materials and Structures \textbf{15}(6), 1513 (2006)

\bibitem{haug1986design}
Haug, E.J., Choi, K.K., Komkov, V.: Design sensitivity analysis of structural
  systems, vol. 177.
\newblock Academic press (1986)

\bibitem{hu1999maximization}
Hu, H.T., Tsai, J.Y.: Maximization of the fundamental frequencies of laminated
  cylindrical shells with respect to fiber orientations.
\newblock Journal of sound and vibration \textbf{225}(4), 723--740 (1999)

\bibitem{hughes2005isogeometric}
Hughes, T.J., Cottrell, J.A., Bazilevs, Y.: Isogeometric analysis: Cad, finite
  elements, nurbs, exact geometry and mesh refinement.
\newblock Computer methods in applied mechanics and engineering
  \textbf{194}(39-41), 4135--4195 (2005)

\bibitem{jihong2006maximization}
Jihong, Z., Weihong, Z.: Maximization of structural natural frequency with
  optimal support layout.
\newblock Structural and Multidisciplinary Optimization \textbf{31}(6),
  462--469 (2006)

\bibitem{liu2017additive}
Liu, C., Du, Z., Zhang, W., Zhu, Y., Guo, X.: Additive manufacturing-oriented
  design of graded lattice structures through explicit topology optimization.
\newblock Journal of Applied Mechanics \textbf{84}(8), 081008 (2017)

\bibitem{liu2019smooth}
Liu, H., Yang, D., Wang, X., Wang, Y., Liu, C., Wang, Z.P.: Smooth size design
  for the natural frequencies of curved timoshenko beams using isogeometric
  analysis.
\newblock Structural and Multidisciplinary Optimization \textbf{59}(4),
  1143--1162 (2019)

\bibitem{meier2014objective}
Meier, C., Popp, A., Wall, W.A.: An objective 3d large deformation finite
  element formulation for geometrically exact curved kirchhoff rods.
\newblock Computer Methods in Applied Mechanics and Engineering \textbf{278},
  445--478 (2014)

\bibitem{nagy2011isogeometric}
Nagy, A.P., Abdalla, M.M., G{\"u}rdal, Z.: Isogeometric design of elastic
  arches for maximum fundamental frequency.
\newblock Structural and Multidisciplinary Optimization \textbf{43}(1),
  135--149 (2011)

\bibitem{olhoff2012optimum}
Olhoff, N., Niu, B., Cheng, G.: Optimum design of band-gap beam structures.
\newblock International Journal of Solids and Structures \textbf{49}(22),
  3158--3169 (2012)

\bibitem{picelli2015evolutionary}
Picelli, R., Vicente, W., Pavanello, R., Xie, Y.: Evolutionary topology
  optimization for natural frequency maximization problems considering
  acoustic--structure interaction.
\newblock Finite Elements in Analysis and Design \textbf{106}, 56--64 (2015)

\bibitem{piegl2012nurbs}
Piegl, L., Tiller, W.: The NURBS book.
\newblock Springer Science \& Business Media (2012)

\bibitem{seyranian1994multiple}
Seyranian, A.P., Lund, E., Olhoff, N.: Multiple eigenvalues in structural
  optimization problems.
\newblock Structural optimization \textbf{8}(4), 207--227 (1994)

\bibitem{shi2017fundamental}
Shi, J.X., Nagano, T., Shimoda, M.: Fundamental frequency maximization of
  orthotropic shells using a free-form optimization method.
\newblock Composite Structures \textbf{170}, 135--145 (2017)

\bibitem{da2017optimization}
da~Silva, G.A.L., Nicoletti, R.: Optimization of natural frequencies of a
  slender beam shaped in a linear combination of its mode shapes.
\newblock Journal of Sound and Vibration \textbf{397}, 92--107 (2017)

\bibitem{simo1985finite}
Simo, J.C.: A finite strain beam formulation. the three-dimensional dynamic
  problem. part i.
\newblock Computer methods in applied mechanics and engineering \textbf{49}(1),
  55--70 (1985)

\bibitem{simo1986three}
Simo, J.C., Vu-Quoc, L.: A three-dimensional finite-strain rod model. part ii:
  Computational aspects.
\newblock Computer methods in applied mechanics and engineering \textbf{58}(1),
  79--116 (1986)

\bibitem{taheri2014simultaneous}
Taheri, A.H., Hassani, B.: Simultaneous isogeometrical shape and material
  design of functionally graded structures for optimal eigenfrequencies.
\newblock Computer Methods in Applied Mechanics and Engineering \textbf{277},
  46--80 (2014)

\bibitem{wang2006maximizing}
Wang, D., Friswell, M., Lei, Y.: Maximizing the natural frequency of a beam
  with an intermediate elastic support.
\newblock Journal of sound and vibration \textbf{291}(3-5), 1229--1238 (2006)

\bibitem{zuo2011fast}
Zuo, W., Xu, T., Zhang, H., Xu, T.: Fast structural optimization with frequency
  constraints by genetic algorithm using adaptive eigenvalue reanalysis
  methods.
\newblock Structural and Multidisciplinary Optimization \textbf{43}(6),
  799--810 (2011)

\end{thebibliography}

%
%

\end{document}